\documentclass[12pt,preprint]{aastex}

\usepackage{graphicx}
\usepackage{amsmath}
\usepackage{mathrsfs}
\usepackage{eucal}

\usepackage{color}

\newcommand{\leftmean}{{\textstyle <}}
\newcommand{\rightmean}{{\textstyle >}}

\newcommand{\lsim}{\raisebox{-3pt}{$\;\stackrel{\small <}{\sim}\;$}}
\newcommand{\gsim}{\raisebox{-3pt}{$\;\stackrel{\small >}{\sim}\;$}} 

\newcommand{\rh}{\text{{\em rh}}} 
\newcommand{\ISR}{S} 

\newcommand{\hourangle}{t_\star}
\newcommand{\celsius}{\,^\circ\text{C}}

\slugcomment{Accepted for publication on the Astrophysical Journal}

\shorttitle{Pressure dependent Habitable Zone}
\shortauthors{Vladilo et al.}

\begin{document}
 
\title{The  habitable zone of Earth-like planets\\ 
with different levels  of atmospheric pressure}

\author{Giovanni Vladilo\altaffilmark{1,2}, 
Giuseppe Murante\altaffilmark{1},  
Laura Silva\altaffilmark{1},\\
Antonello Provenzale\altaffilmark{3}, 
Gaia Ferri\altaffilmark{2},
Gregorio Ragazzini\altaffilmark{2}
}
%
%
%
%
%

\email{vladilo@oats.inaf.it}

\altaffiltext{1}{INAF - Trieste Astronomical Observatory, Trieste, Italy}
\altaffiltext{2}{Department of Physics, University of Trieste, Trieste, Italy}
\altaffiltext{3}{Institute of Atmospheric Sciences and Climate - CNR, Torino, Italy}

%
%

\begin{abstract}
As a contribution to the study of the habitability of extrasolar planets,
we implemented  
a 1-D Energy Balance  Model (EBM), 
the simplest seasonal model of planetary climate,
with new prescriptions for most physical quantities. 
%
Here we apply our EBM to investigate the surface habitability of planets with an Earth-like atmospheric composition but different levels of surface pressure. The habitability,  
defined as the mean fraction of the planet's surface on which liquid water could exist,  is
estimated from the pressure-dependent liquid water temperature range, taking into account seasonal and latitudinal variations of surface temperature. By running several thousands of EBM simulations we  generated a map of the habitable zone (HZ) in the plane of the orbital semi-major axis, $a$, and surface pressure, $p$, for planets in circular orbits around a Sun-like star. As pressure increases, the HZ becomes broader, with an increase of 0.25\,AU in its radial extent from $p$=1/3\,bar to $p$=3\,bar. At low pressure, the habitability is low and varies with $a$; at high pressure, the  habitability is high and relatively constant inside the HZ.  We interpret these results in terms of the pressure dependence of the greenhouse effect, the efficiency of horizontal heat transport, and the extent of the liquid water temperature range. Within the limits discussed in the paper, the results can be extended to planets in eccentric orbits around non-solar type stars. The main characteristics of the pressure-dependent HZ are modestly affected by variations of planetary properties, particularly at high pressure. 
%
\end{abstract}


\section{Introduction}

Observational searches for extrasolar planets are motivated,
in large part, by the quest for astronomical environments
with physical and chemical conditions supportive of life.
The criterion most commonly adopted to define such 
``habitable'' environments is the presence of water in liquid phase. 
This criterion is motivated by the fundamental role played by water in terrestrial life
and by the unique properties of the water molecule (Bartik et al. 2011). 
%
%
%
Among all types of astronomical environments, 
only planets and moons may possess the right combination of temperature
and pressure  
compatible with water in the liquid phase.  
The exact range of planetary physical conditions is determined
by a number of stellar, orbital and planetary factors.
The combination of stellar flux and orbital parameters that yield
surface planet temperatures 
compatible with the  liquid water criterion 
defines the circumstellar ``habitable zone'' 
 (Dole 1964, Hart 1979, Kasting et al. 1993). 
The  location of the inner and outer boundaries of the  habitable zone (HZ)
depends on many planetary factors and, in particular, 
on the atmospheric properties that govern the greenhouse effect. 
The outer limit of the ``classic'' HZ
is  calculated 
allowing for the presence of a geochemical cycle
of CO$_2$ that
creates a stabilizing climate feedback;
the inner limit takes into account the possibility of a runaway
greenhouse effect driven by water vapor 
(Walker et al., 1981; Kasting et al., 1993; Kasting \& Catling, 2003; Selsis et al. 2007). 
Planets  with high-pressure,  H$_2$-He atmospheres  
would be habitable well outside the outer edge  of the classic HZ
(Pierrehumbert \& Gaidos, 2011).  
In the context of HZ studies, 
the possible existence of habitable exomoons 
is also under investigation   
(Reynolds et al. 1987, Williams et al. 1997, Scharf 2006, Heller \& Barnes 2013).

The  HZ
concept was introduced in  scientific
literature before the first discovery of an
extrasolar planet with the radial-velocity method (Mayor \& Queloz, 1995).
The subsequent detection of hundreds of exoplanets with the same method
and/or with the transit method has converted the HZ concept
into a powerful tool used to discriminate habitable planets on the basis
of the orbital semi-major axis, a physical quantity that can be derived
from both  detection methods. 
One of the main results of exoplanet observations is
the discovery of a great variety of planetary and orbital characteristics
not found  in the Solar System
(see e.g. Udry \& Santos, 2007,
Howard et al., 2012, and references therein).
Even if a large region in this parameter space yields
conditions not appropriate for liquid water,
a fraction of habitable planets are expected to be present.
At the present time, the number of planets detected inside or close to the HZ
is  small (Selsis et al. 2007, Pepe et al. 2011, Borucki et al. 2012, 
Anglada-Escud\'e et al. 2012, Tuomi et al. 2013),
but this number is expected to increase dramatically in the coming years. 
In fact,
the number of low-mass, terrestrial planets potentially in the HZ
is expected to be very high since the planetary initial mass function  peaks
at low masses (Mordasini et al.  2012)
%
%
and the multiplicity of planetary systems is higher when
low-mass planets are detected (Lo Curto et al. 2010,
Lissauer et al. 2011, Latham et al. 2011). 
Exploratory studies  of terrestrial planets in the HZ
will set the framework for focusing subsequent, time-consuming
investigations aimed at the search for atmospheric biomarkers.

The measurement of the physical quantities relevant for  habitability 
suffers from the limitations inherent to the 
observational techniques of exoplanets (Udry \& Santos, 2007). 
%
Given the shortage of experimental data on terrestrial-type exoplanets,
the study of their habitability requires a significant effort of modelization.
Models of planetary climate are fundamental in this context, since they
complement the observational data with quantitative predictions
of the  physical quantities relevant for assessing their habitability. 

A  variety of
models of planetary climate are currently available, all originated
from studies of the Earth's climate (McGuffie \& Henderson-Sellers, 2005). 
State-of-the art, global circulation models (GCMs) 
allow us to treat in 3-D the chemistry  and dynamics of the atmosphere, as well as to track  the 
feedbacks  existing between the different components of the climate system. 
%
The use of 3-D climate models to investigate
the habitability of extrasolar planets is quite recent.
So far, this technique has been applied to a few planets
(or candidate planets) 
orbiting M dwarf stars 
(Joshi 2003, Heng \& Vogt 2011, Wordsworth et al. 2011).
%
Modeling the climate 
requires a large
number of planetary parameters not constrained by observations of exoplanets.
Given the very large amount of computing resources required to run a GCM,
the exploration of the parameter space relevant to the climate
and habitability requires a more flexible tool.  

%
%
%
%
%
%
%
%

Energy Balance Models (EBM) offer an alternative approach to  climate modelization. 
These models employ simplified recipes for the physical quantities relevant
to the climate and  require a modest amount of CPU time and 
 a relatively low number of input parameters.
The predictive power is  limited since EBMs
do not consider, among other effects, the wavelength dependence of the radiative transfer
and the vertical stratification of the atmosphere.
In spite of these limitations, EBMs offer the possibility to
estimate  the surface temperature 
at different latitudes and seasons, 
and are ideal for exploratory studies of habitability.
Feedback processes, such as the ice-albedo feedback 
(Spiegel et al. 2008; hereafter SMS08)
or the CO$_2$ weathering cycle (Williams \& Kasting 1997; hereafter WK97)
can be implemented,  although in a schematic form.

%
%

Previous applications of EBMs to
extrasolar planets have investigated the dependence of
the habitability on axis obliquity, continent distribution, 
CO$_2$ partial pressure, rotation period
and  orbital eccentricity
(WK97, Williams \& Pollard 2002, SMS08, Spiegel et al. 2009, Dressing et al. 2010).
Climate EBMs have also been used 
to explore the habitability
in the presence of Milankovitch-type cycles (Spiegel et al., 2010),
 in tidally locked exoplanets (Kite et al. 2011),
and around binary stellar systems (Forgan, 2012).
Predictions of planet  IR light curves can also be obtained with EBMs
(Gaidos \& Williams 2004).
Here we introduce a more complete formulation of a planetary EBM,
aimed at addressing open conceptual questions in planetary habitability and paleo-climate dynamics. 
As a first application of this model, in the present paper we
investigate the influence of atmospheric pressure on 
 planet temperature and habitability.
The focus is on the physical effects induced by variations of the total surface
pressure, $p$, at a constant chemical composition of the atmosphere. 
Surface pressure is
a key thermodynamical quantity required to estimate the 
habitability via the liquid water criterion. At the same time,
pressure influences the climate  in different ways 
and the high computational efficiency of EBMs allows us to explore 
pressure effects under a variety of initial conditions.
%
%
%
Here we have made an intensive use of EBM  
to generate maps of planetary habitability as a function of $p$ and
semi-major axis, i.e. a sort of pressure-dependent HZ.
The calculations have been repeated for several combinations
of orbital and planetary parameters.
In particular,
we have explored how the climate and habitability
are affected by changes of physical quantities that  are  
not measurable with present-day exoplanet observations. 

The  paper is organized as follows. 
The climate model is presented in the next section. 
Technical details on the model prescriptions and calibration are given in  the appendix 
(Section \ref{modelPrescriptions}). 
In Section 3 we present the habitability maps obtained from our simulations.
The results are discussed in Section 4 and  the work summarized in Section 5.

\section{The climate model \label{sectModel}}

The simplest way of modeling the climate  of a planet
is in terms of the energy balance between the incoming and outgoing radiation.
%
The incoming radiation, $\ISR$, is of stellar origin and peaks in the visible, with
variable contributions in the UV and near-IR range, depending on the spectral type
of the central star. 
The outgoing radiation emitted by the planet, $I$, generally peaks
at longer wavelengths and is called {\em outgoing long-wavelength radiation} (OLR).
For the planets that can host life, characterized by a surface temperature $T \approx 3 \times 10^2$\,K,
the OLR peaks in the thermal infrared.
%
In addition, the planet reflects back to space
a fraction $A$ of the short-wavelength stellar radiation. This fraction, called {\em albedo}, 
does not  contribute to the heating of the planet surface.
At the zero order approximation we 
require that the fraction of stellar radiation absorbed by the planet, $S(1-A)$,
is balanced, in the long term, by the outgoing infrared radiation, i.e. 
$I = \ISR(1-A)$.

The zero-order energy balance neglects the  horizontal transport, i.e the exchanges of heat
along the planet surface. 
Energy balance models (EBMs) provide a simple way to include the horizontal transport 
in the treatment of planetary climate.  
In EBMs, the planet surface is divided in strips delimited by latitude circles,
called ``zones'', and the physical quantities of interest are averaged
in each zone over one rotation period. 
The longitudinal heat transport does not need to be explicitly considered
since  it is averaged in each zone. 
The treatment of the horizontal transport is thus restricted
to that of the latitudinal transport.

Zonally averaged EBMs are one-dimensional 
in the sense that the spatial dependence of the physical quantities 
only takes into account the latitude, $\varphi$,
usually mapped as $x=\sin \varphi$.
However, with
the inclusion of a term describing the effective thermal capacity
of the planet surface, one can also introduce the
dependence  on time, $t$.
At each given time and latitude zone
the thermal state of the atmosphere and ocean 
is represented by a single temperature, $T=T(t,x)$, representative of the surface temperature. 
This is the type of model that we consider here.
In particular, 
following previous work  on Earth and exoplanet climate
(North \& Coakley 1979, 
North et al. 1983,
WK97, SMS08),
we adopt the {\em diffusion equation} of  energy balance  
\begin{equation}
C {\partial T \over \partial t} - 
{\partial \over \partial x}
\left[ D \, (1-x^2) \, { \partial T \over \partial x} \right]
+ I = S \, (1-A) ~~~.
\label{diffusionEq}
\end{equation}
In this equation, 
the efficiency of the latitudinal heat transport is governed
by the diffusion coefficient, $D$, while the thermal inertia
of the different components of the climate system is determined
by the effective heat capacity, $C$.
%
%
The incoming short wavelength radiation, $S$, is an external forcing driven
by astronomical parameters, such as the stellar luminosity, the orbital
semi-major axis and eccentricity,
and the obliquity of the planet axis of rotation. 
The outgoing infrared radiation, $I$, is largely governed 
by the physical and chemical properties of the atmosphere. 
The albedo $A$ is specified by the surface distribution of continents, oceans, ice, and clouds.
The physical quantities generally evolve with time, $t$.
The temporal dependence of EBMs can be used to
study seasonal and/or long-term climate effects.  
In this work we only consider the seasonal evolution. 
At variance with the zero dimensional model, 
the physical quantities in the EBM are zonal, i.e. they depend on $x=\sin \varphi$.
The dependence of the physical quantities on $t$ and $x$  
may be indirect, via their dependence on the temperature $T=T(t,x)$. 
Some quantities, such as the albedo of the oceans, depend on the
stellar zenith distance, $Z_\star=Z_\star(t,x)$. 
In spite of its simplicity, the model is able to capture
some of the feedback between the main components of the climate system,
such as the ice-albedo feedback.

\subsection{Physical ingredients \label{sectIngredients}}

%
%

The degree of realism of the EBM is largely determined
by the  formalism adopted to describe 
the physical quantities $C$, $D$, $I$, $S$, and $A$.
The recipes that we adopt 
are inspired by previous EBM work published
by SMS08, WK97, and North et al. (1983).
In our EBM, 
we have introduced several new features
aimed at addressing some limitations of the method
and broadening the range of application of the model. 

An intrinsic limitation of diffusive EBMs is 
that the physics that drives the horizontal transport is
more complex than implied by the diffusion term  
that appears in Eq.\,(\ref{diffusionEq}).
We have introduced
a dependence of the diffusion coefficient $D$ on the
stellar zenith distance. 
In this way, the efficiency of horizontal heat transport becomes a
function of latitude and orbital phase, as in the case of real planets.
In a real planetary atmosphere, heat transport across latitudes is governed by large-scale fluid motions and their instabilities (such as baroclinic instability). Here, such effects are parametrized in a overly simplified way by assuming diffusive transport, albeit with a variable diffusion coefficient.

Previous EBM work has been applied to planets with Earth atmospheric pressure.
Given the impact f atmospheric pressure on the climate and habitability, 
we have generalized the model for applications
to terrestrial planets with different levels of surface pressure.
With this aim, in addition to the pressure dependence of the diffusion coefficient, $D$,
already considered by WK97, we have also taken into account 
the pressure dependence of the OLR, $I$,
and of the effective thermal capacity, $C$. 
Our model allows   permanent ice covers to be formed
and calculates the cloud coverage according to  the type of underlying surface.
All the above features are presented in the appendix,
where we provide a full description of the method.

\subsection{Solving the diffusion equation}

The diffusion equation (\ref{diffusionEq}) is a partial differential equation in two variables, $x$ and $t$,
that can be solved by separating the terms containing the temporal and spatial derivates.  
The temporal part can be  written in the form $\partial T /\partial t = f(t,T[x])$, which can be solved
with the Runge-Kutta method.
In practice, we use the routine {\em odeint} (Press et al. 1992) 
characterized by an adaptive stepsize control. 
To solve the spatial part, $f(t,T[x])$, we use the Euler method.   
To this end, we discretize
$x$ in $N$ equispaced cells, delimited by $(N+1)$ cell borders, and use a {\em staggered grid}, i.e.
the quantities of interest are calculated at the center of the cells, whereas their derivates 
are calculated at the borders.
Boundary values are obtained assuming that the diffusion term and the partial derivative of $T$ are null at the poles. 

To validate the code we used the same set of prescriptions of the physical quantities
adopted by  SMS08 and we verified 
the capability of the code to reproduce the results published in that paper. 
As an example, we show in Fig. \ref{earthLatTemp} the mean annual temperature-latitude
profile of the Earth obtained in this way (dotted line), which is the same shown in Fig. 2 of SMS08 (solid line). 
We also verified that our code recovers
the exact same  
seasonal evolution of the zonal temperatures
(Fig.\,3 in SMS08), as well as the ``snowball'' transition predicted to occur when
the latitudinal diffusion is decreased by a factor of 9 (Fig. 4 in SMS08). 


\subsection{Model calibration\label{modelCalibration}}

\begin{figure}[ht]    
\begin{center} 
\includegraphics[width=8cm]{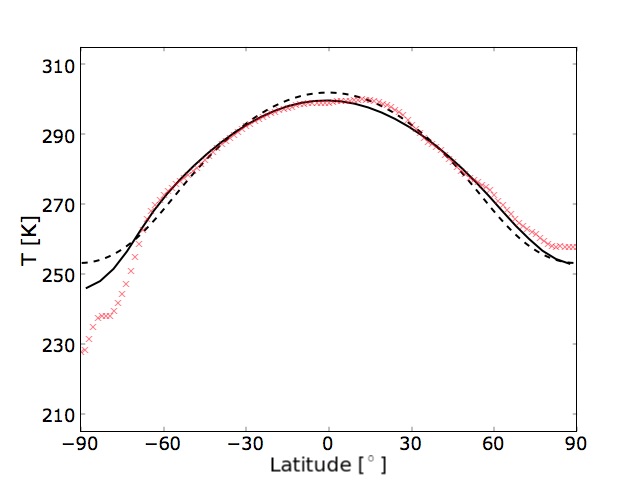}
\includegraphics[width=8cm]{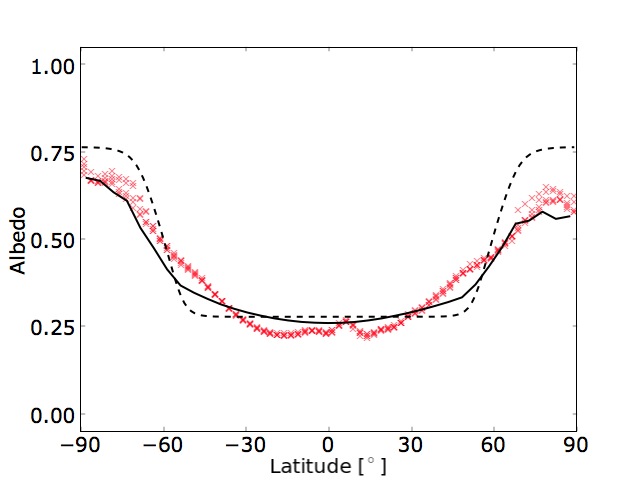} 
\caption{
Comparison of experimental data and model predictions of
the Earth latitude profiles of   
mean annual temperature (left panel) and mean annual albedo (right panel). 
Crosses:  average ERA Interim temperatures
in the period 1979-2010 (left panel);
ERBE short-wavelength albedo in the years 1985-1989 (Pierrehumbert 2010).
Solid line: our model. 
Dashed line: model by SMS08.  
  }
\label{earthLatTemp}%
\end{center}
\end{figure}

\begin{figure}[ht]   
\begin{center}
\includegraphics[width=8cm]{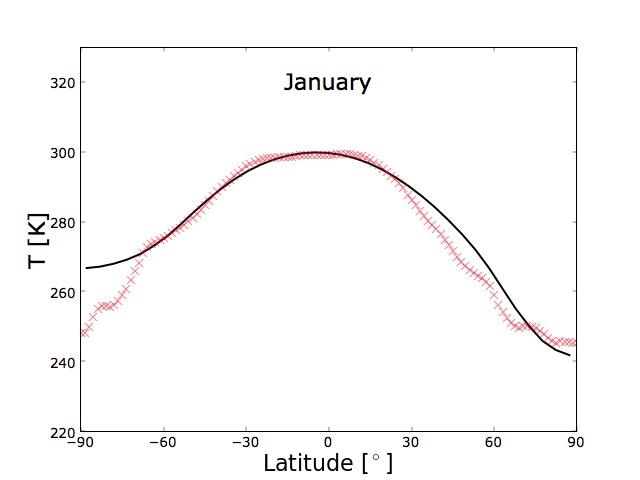}
\includegraphics[width=8cm]{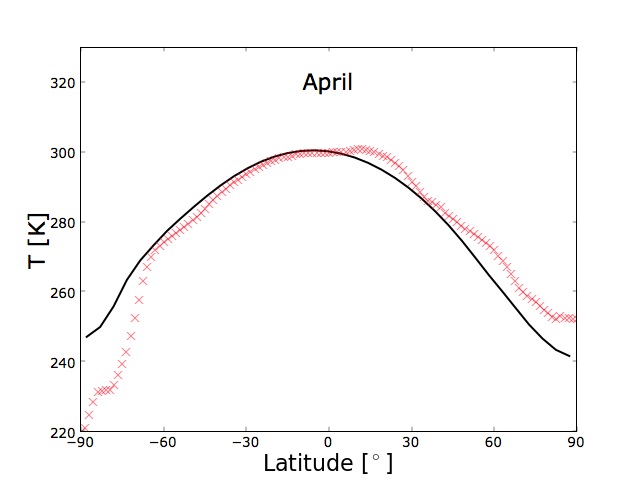} 
\includegraphics[width=8cm]{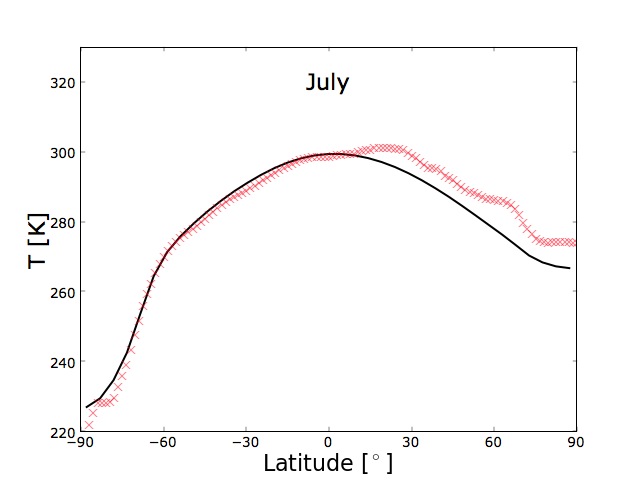} 
\includegraphics[width=8cm]{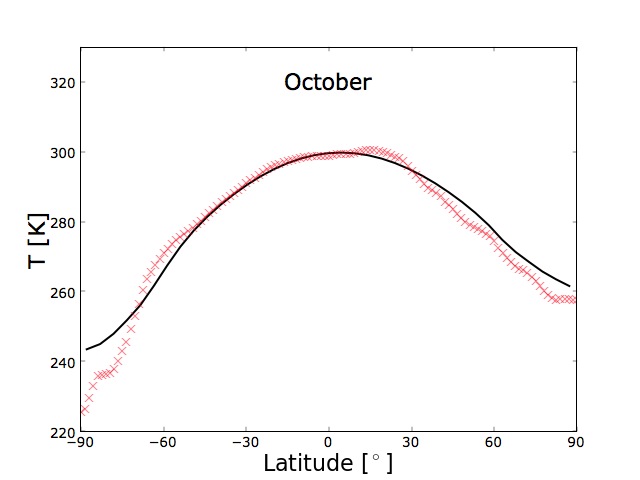} 
\caption{ 
Comparison of experimental data and model predictions of the Earth
latitude profiles of mean monthly temperature for January, April, July and October. 
Crosses: average ERA Interim data for the same months collected in the period 1979-2010.
Solid line: predictions of our EBM.
  }
\label{earthSeasons}%
\end{center}
\end{figure}

After the adoption of our own set of physical recipes,
described in the appendix,
the model was calibrated to reproduce the Earth climate data. 
%
The calibration was performed 
by adopting astronomical and planetary parameters appropriate
for the Earth and tuning the remaining parameters 
in such a way to reproduce 
the surface temperature and albedo 
of the Earth (Figs. \ref{earthLatTemp} and \ref{earthSeasons}).
%

%

The set of Earth data adopted for the calibration
is listed in Table \ref{tabEarthData}.  
%
The incoming stellar radiation,
$S=S(t,x)$, 
is fully specified 
by the solar constant and the Earth orbital parameters and obliquity. 
The zonal values of Earth ocean fraction $f_o$,
not shown in the table, were taken from Table III of WK97.  

The set of fiducial parameters that we adopt for our EBM is shown in 
Table \ref{tabCalibratedPar}. 
Some of these parameters are taken from previous work,
cited in the last column of the table. Further information on
the choice of the  parameters is given in the appendix. 

After the calibration procedure, our Earth model yields 
a mean annual global temperature
$T_{m,\circ} = 287.8$\,K and 
a mean annual global albedo
$A_{m,\circ} = 0.320$,
in fine agreement with the Earth 
values (Table \ref{tabEarthData}).
%
%
In the left panel of Fig. \ref{earthLatTemp},
we compare the mean annual temperature-latitude profile predicted by our model
(solid line) with experimental data (crosses). 
The agreement with the observations is excellent in most latitudes,
with an area-weighted rms deviation of 2.2\,K. However,
the model fails to reproduce adequately the temperature profile
at Antarctic latitudes. 
This is likely due to the peculiarities of Antarctica,
such as its high elevation with respect to the sea level,
that are not accounted for in  EBMs.
In any case, compared to previous EBM work, our model yields a better agreement
at Antarctic latitudes. 
The predictions of the model of SMS08 are shown for comparison   in  Fig. \ref{earthLatTemp}
(dashed line);
the model by WK97 predicts a mean annual temperature of 260 K 
at the southernmost latitudes (see their Fig. 4),
well above the experimental data. 
%
%
%
 
%
In the right panel of Fig. \ref{earthLatTemp} we compare the mean annual 
albedo-latitude profile of the Earth 
predicted by our EBM  (solid line) with the  experimental data
obtained by ERBE for the period 1985-1989 (crosses). 
The agreement is reasonable but not perfect.
The main reasons for disagreement are:
(i) the over-simplified physical
prescriptions of the EBM, which do not consider 
the atmospheric scattering of the short-wavelength radiation;
(ii)
the limited amount of fine tuning
of the albedo parameters (e.g. a constant value of albedo for all  land).
For the sake of comparison, in the  same figure 
we show the mean annual albedo profile obtained
with the Earth model of SMS08 (dashed line), where the albedo
is a simple function of  temperature. 
At variance with this previous formulation, our parameters can be varied to model the albedo of
planets with different types of surfaces. 
It is worth mentioning that the top-of-atmosphere recipes for the albedo
given by WK97, which include a treatment of atmospheric scattering, 
cannot be applied to the present work. This is because the radiative-convective
calculations of WK97 were performed by varying the concentration of CO$_2$,
while we are interested in varying the total pressure
keeping  the composition fixed.  
 
Careful inspection of the  seasonal dynamics 
shows that in the northern hemisphere the temporal
evolution of the model slightly lags that of the data (see Fig. \ref{earthSeasons}).
This is probably due to an intrinsic limitation of 1-D EBMs, 
related to the calculation of the zonal thermal 
capacity:
 the zonal average of $C$ tends to be overestimated  
in zones with ocean fractions above $\approx 0.1$ 
because  the effective thermal capacity 
of the oceans is much higher than that of the lands (WK97).
As a consequence,
our EBM overestimates the thermal inertia at the mid-northern latitudes of the Earth
(with ocean fractions $f_o = 0.3-0.5$) and predicts that the seasonal temperature should vary more
slowly than is observed.
%

%
For the purpose of this investigation, an important feature of the temperature profile
is the location of the ice line, which affects the calculation of planetary habitability.
To test this feature, we compared the latitude $\varphi_\circ$ where $T=273\,$K 
in our EBM and in the Earth data.
We find a mean annual difference 
$\leftmean\varphi_{\circ,\text{model}}-\varphi_{\circ,\text{data}}\rightmean
=+0.6^\circ \pm 3.2^\circ$ in the southern hemisphere
and $-2.4^\circ \pm 10.4^\circ$ in the northern hemisphere
(average over 12 months).
The small offset $\leftmean\varphi_{\circ,\text{model}}-\varphi_{\circ,\text{data}}\rightmean$
in both hemispheres testifies in favour of the quality of the ice line calibration. 
The large scatter in the northern hemisphere is due to the above-discussed
seasonal time lag resulting from the  overestimate of the thermal capacity.  

More details on the model calibration are given in the Appendix,
where we show that, in addition to the temperature and albedo data, also the OLR data 
of the Earth have
been used to constrain our EBM (see Section \ref{sectOLRcal}).

\section{Applications to exoplanets \label{sectResults}} 

The climate model 
presented in the previous section is a flexible tool that
allows us to explore part of the huge parameter space 
relevant for the study of planetary habitability. 
The physical quantities that enter into the EBM are quite numerous 
and it is important to make a distinction between those that are
observable and those that are not  in the case of exoplanets.
In the present work we generate maps of habitability
in a 2-D  space built-up in such a way to explore how a parameter
unconstrained by observations influences the habitability 
while we vary an observable physical quantity.
In this way, we 
investigate the effects of unconstrained parameters 
against an experimental parameter space where one can 
place observed exoplanets. 

The choice of the observational quantities that we use in our maps
is dictated by their relevance to the climate.
Luckily, at least some of the model parameters most relevant to the  planet's 
climate
can be constrained by  observations. This is true for the
stellar flux, the semi-major axis and the eccentricity of the planet orbit.
These three parameters can be combined to derive 
the mean annual level of insolation, a fundamental parameter for the climate. 
To define the HZ in our maps we will show the
correspondence between this quantity and the semi-major axis 
for the case of planets orbiting a solar-like star. 

%
Among the quantities unconstrained by observations,
the axis obliquity and CO$_2$ partial pressure
have been investigated in previous work (WK97, Spiegel et al. 2009).
Here 
we focus on the surface pressure, $p$, in planets with
Earth-like atmospheres.  
Our maps of habitability  therefore have  the semi-major axis or insolation in the abscissae
and the surface pressure in the ordinates. 
In this section, we explain how we generate this type of map.
In the next section we investigate how the results that we obtain
are influenced by other unconstrained planetary parameters, such as 
axis obliquity, rotation period, geography and albedo. 

\subsection{Running the simulations\label{running}}

To investigate the effects of surface pressure on the location of the HZ
we generate a large number of models of an Earth-like planet by varying  
$a$ and   $p$
while keeping constant the other orbital and planetary parameters.
To focus our problem we consider an atmosphere with an Earth-like composition
but variable level of total pressure. In practice,
we scale  the partial pressures of non-condensable gases by a constant factor. 
In this way, the mixing ratios of non-condensable greenhouses,
such as CO$_2$ or CH$_4$, are always equal to that of the Earth atmosphere.
The mixing ratio of water vapor is instead set by the temperature and relative humidity.
This scaling of the pressure is in line with the assumptions adopted
in the calibration of the OLR 
described in Section \ref{sectOLRcal}.

Each simulation
starts with an assigned  value of temperature  in  each zone, $T_\text{start}$, 
and is stopped when the mean global annual temperature,
$T_m$, converges within a prefixed accuracy. 
In practice, we calculate the increment $\delta T_m$
every 
10 orbits and stop the simulation when  
$| \delta T_m | < 0.01$\,K.
The convergence is generally achieved in fewer than 100 orbits.
Tests performed with widely different values of $T_\text{start}> 273$\,K
indicate that the simulations converge to the same solution.
We adopted a value of $T_\text{start}=275$\,K, just above the threshold
for ice formation, so that the simulation starts without ice coverage
and without artificial triggering of snowball episodes. 
The choice of a low value of $T_\text{start}$ 
is dictated by our interest in exploring a wide range of pressures:
$T_\text{start}=275$\,K is  below the water boiling point
even at very low values of pressure, when the boiling point is just a few kelvins above the freezing point.

%

%

The starting value of  total pressure $p$ in the simulations is
the total pressure of non-condensable gases, i.e. the ``dry air'' pressure. 
In the course of the simulation, $p$ is updated
by adding the current value of the partial pressure of water vapor.
This is  
calculated as
\begin{equation}
p\text{H}_2\text{O} = \rh \times p_{\text{sat},w}(T) ~~~,
\label{pWater}
\end{equation}
where 
$\rh$ is the relative humidity and $p_{\text{sat},w}(T)$ is the 
saturated water vapor pressure\footnote{
We adopt 
$p_{\text{sat},w}(T)=e^{(77.3450+0.0057 \, T-7235/T)}/T^{8.2}$,
a relation 
that yields an excellent agreement with the values tabulated 
in the CRC Handbook of Chemistry and Physics (Lide 1997). 
See also http://www.engineeringtoolbox.com/water-vapor-saturation-pressure-air-d\_689.html. 
}.
Also the specific heat capacity, $c_\text{p}$, and the mean molecular weight, $m$, 
of the atmosphere are recalculated at each orbit
taking into account the contribution of water vapor. The thermal capacity
and the diffusion coefficient, which depend on  $c_\text{p}$ and  $m$, are also updated. 

In addition to the regular exit  based
on the convergence criterion, the simulations are  stopped
when the mean planet temperature 
exceeds some prefixed limits $(T_\mathrm{min},T_\mathrm{max})$.
This forced exit without convergence is introduced 
to minimize the computing time when we run a large number
of simulations, as in our case.
We adopt $T_\mathrm{min}=220$\,K,
well below the limit of liquid water habitability. 
The value of $T_\mathrm{max}$ is based on the 
 water loss limit  criterion explained below.
When $T_m$ is outside the interval $(T_\mathrm{min},T_\mathrm{max})$,
the simulation is stopped and the indices of habitability are set to zero. 

%

The existence of a planetary water loss limit is based on the following arguments. 
At a given  value of relative humidity, 
$p$H$_2$O increases with temperature
according to Eq. (\ref{pWater})
because the saturated pressure of water vapor increases with $T$. 
The dominant feedback mechanism of water vapor is the enhancement of
the IR opacity, which tends to raise the temperature.
In extreme cases, this positive feedback may lead to 
a complete water vaporization via a runaway greenhouse effect,
followed by a loss of hydrogen in the upper atmosphere
via EUV stellar radiation.  
All together, these effects indicate the existence of a temperature limit
above which water is  lost from the planet.
In the case of the Earth, the water loss limit
is predicted to occur at $\simeq 340$\,K, while for a planet with 
$p=5$\,bar at $\simeq 373$\,K  (Selsis et al. 2007 and refs. therein). 
These values of temperature approximately lie  at  90\% 
of the liquid water temperature range calculated at the corresponding value of pressure.
On the basis of these arguments, we adopt as a water loss limit the value
\begin{equation}
T_\mathrm{max}= T_\text{ice}(p) + 0.9 \times \left[ T_\text{vapor}(p) - T_\text{ice}(p) \right]
\label{eqTmax}
\end{equation}
where $T_\text{ice}(p)$ and $T_\text{vapor}(p) $ are
the melting and boiling points of water at  pressure $p$ (Lide 1997). 
The adoption of the water loss limit (\ref{eqTmax})
minimizes the difficulty of tracking the effects
of water vapor in the atmosphere 
when the temperature is high (see Section \ref{sectOLRcal}).
The results on  runaway water-loss events are conservative given our choice of adopting
a ``cold start'' in the simulations. 

%
%

%


\subsection{Indices of mean temperature and habitability\label{sectHabIndex}}


For each  set of input parameters, our  simulations
provide a matrix of surface temperatures calculated
at discrete values of latitude and time, $T(\varphi,t)$. 
We use different tools to analyze these results.
To cast light on specific cases, we  investigate the
latitude-temperature profile and its seasonal evolution.
To have a global view of a large set of results, we build up maps  
of mean annual global temperature and habitability.




We call {\em mean planet temperature}, $T_m$, the mean global annual temperature
at the planet surface. 
To calculate this quantity, we average $T(\varphi,t)$ 
on latitude and time. The average on latitude is  weighted in area. 


%
To estimate the surface habitability from the  results of our simulations,  we require
that the temperature 
and the pressure 
lie within the range of liquid water in the phase diagram of H$_2$O. 
We consider values of total pressure above the triple point and below the critical pressure.
Within this interval,
we define a {\em habitability function}  such that
 \begin{equation} 
H(\varphi,t) = 
\begin{cases}
1 &  \text{if $T_\mathrm{ice}(p) \leq T(\varphi,t) \leq T_\mathrm{vapor}(p)$} \\
0 &  \text{otherwise} ~~~.
\end{cases}
\label{habitabilityFunction}
\end{equation}

Several indices of surface habitability can be defined by integrating $H(\varphi,t)$  in different ways. 
Following SMS08, we
introduce the temporal habitability
\begin{equation}
f_\mathrm{time}(\varphi) =  {\int_{0}^P dt ~ H(\varphi,t)  \over P} ~~~,
\label{f_time}
\end{equation}
 i.e. the fraction of orbital period $P$ in which the planet is habitable at a given latitude,
and  the regional habitability
\begin{equation}
f_\mathrm{area}(t)= 
\frac{\int_{-{\pi \over 2}}^{+{\pi\over 2}} \, d\varphi \, 
\left[ H(\varphi,t) \, \cos \varphi \right]}{2} ~~~,
\label{f_area}
\end{equation}
i.e. the fraction of planet surface that is habitable at a given time. 
By integrating the habitability function
both in latitude and time, we obtain the mean global annual habitability
\begin{equation}
h = 
\frac{\int_{-{\pi\over 2}}^{+{\pi \over 2}}  \, d\varphi \, \int_{0}^P \, dt \, 
\left[ H(\varphi,t)  \, \cos \varphi \right] }{2P}   ~.
\label{h_index}
\end{equation}
This represents the {\em mean fraction of planet surface that 
is habitable during the orbital period}.

In addition to these indices, already defined by SMS08, we introduce here the
index of continuous habitability
\begin{equation}
h_c = 
\frac{\int_{-{\pi\over 2}}^{+{\pi \over 2}}  \, d\varphi \, f'(\varphi)  
\, \cos \varphi  }{2}   ~,
\label{defHabconOne}
\end{equation}
with
 \begin{equation} 
f'(\varphi) = 
\begin{cases}
1 &  \text{if $f_\mathrm{time}(\varphi) =1$} \\
0 &  \text{if $f_\mathrm{time}(\varphi) <1$} ~.
\end{cases}
\label{defHabconTwo}
\end{equation}
The index $h_c$ represents the {\em fraction of planet surface that is continuously habitable
during the orbital period}. This index vanishes
if all the latitude zones undergo a period of non habitability in the course of the orbital period.
By construction, it is always $h_c \leq h$. 


\subsection{Habitability maps \label{maps}}

In Fig. \ref{map_Thap_ptot} we show the map of habitability
obtained from our simulations for an Earth-like planet in circular orbit
around a Sun-like star. 
To obtain this figure, we run
a total  of 4032 simulations 
covering the interval of semi-major axis $0.65 \text{\,AU} \leq a  \leq 1.35 \text{\,AU}$,
with a step $\delta a = 0.01$\,AU,
and the  interval  of  pressure 
$10^{-2.0} \text{\,bar}\leq p  \leq 10^{+0.8} \text{\,bar}$,
with a constant logarithmic step $\delta \log p \text{\,(bar)} = 0.05$. 
The values of the  parameters that are  
kept constant in these simulations are listed in Table \ref{multirunPar}.

The map 
of Fig. \ref{map_Thap_ptot} shows
the results obtained at each point of the plane ($a$, $p$)
in terms of mean annual global temperature and habitability.
The filled circles in the map indicate the positions in the plane ($a$, $p$)
where $h>0$;
%
the value of total pressure associated to these symbols 
includes the partial pressure of water vapor
updated in the course of the simulation. 
%
Crosses 
indicate positions on the plane where the simulations were forced to exit;
the total pressure for these cases is
the starting value of dry air pressure (see Section \ref{running}). 
Empty areas of the map, as well as crosses, indicate a location of non-habitability.

The  simulations yield information not only on the degree of habitability,
through the index $h$, but also on the ``quality'' of the habitability, 
through a detailed analysis of the seasonal and latitudinal variations of the temperature,
as we shall see in the next section. However, some cautionary remarks must be made
before interpreting these data.


The model has been calibrated using Earth climatological data.
These data span a  range of temperatures, roughly
220\,K $\lsim T \lsim$ 310\,K,  not sufficient to cover
the broad diversity expected for  exoplanets,
even if we just consider  those of terrestrial type. Given the fundamental role
of temperature in the diffusion equation, one should be careful
in using the physical quantities outside this range, where direct calibration is not possible.
In this respect, the major reason of concern is the estimate  
of the outgoing long-wavelength radiation that has been done with
radiative calculations (see Section \ref{sectOLRcal}). The difficulty of calibrating the OLR
outside the range of terrestrial temperatures 
makes uncertain the exact localization of the inner and outer edges of the HZ.
In particular, the results with $T_m \gsim 330$\,K 
should be treated with caution,
given the strong effects of water vapor predicted to occur in this
temperature range, which are not directly testable.  
These cases lie in the region of high pressure in Fig. \ref{map_Thap_ptot}
(symbols color-coded in orange and red). 
The fact that in these cases also the pressure is quite different from 
the Earth value makes these results particularly uncertain.
In the following discussion, we will consider these 
particular results to be purely tentative. 
We note that the difficulty of making climate predictions outside
the parameter space sampled by the Earth is a common problem
of any type of climate model, no matter how sophisticated. 
In this respect, simple models, like our EBM,  help to obtain 
preliminary predictions to be tested by subsequent investigations. 




\section{Discussion \label{sectDiscussion}}

In this section we  describe and interpret the complex
patterns that we find in the pressure-dependent map of 
planet temperature and habitability of  Fig. \ref{map_Thap_ptot}.
We then discuss how the results can be extended to more general situations
other than circular orbits of planets orbiting a Sun-like star. 
We conclude this section setting our results in the context of previous studies.

\subsection{The pressure-dependent habitable zone}


The circumstellar HZ  shown  in Fig. \ref{map_Thap_ptot} shows several characteristics,
in terms of mean planet temperature and habitability, that can be summarized as follows.

The radial extent of the HZ increases with $p$. 
The outer edge extends from 1.02 to 1.18 AU when the pressure  rises from
0.1  to 3 bars.
The inner edge approaches the star from 0.87 to 0.77 AU in the same pressure interval.
No habitability is found below $p \simeq  15$\,mbar.  

The broadening of the HZ with increasing pressure
is accompanied by an increase of the interval of mean planet temperatures
spanned at constant $p$.
At high pressures most of the broadening of the HZ  
is contributed by  the area of the plane where the solutions have mean
temperatures $T_m \gsim 60\celsius$ (i.e. $T_m \gsim 333\,$K; 
orange and red symbols above the red line in the figure). 
If we focus on the  interval of mean temperatures  
$0 \celsius \lsim T_m \lsim 60 \celsius$  
(region with $273\,\text{K} \lsim T_m \lsim 333 \,\text{K}$
between the magenta and red line), 
the broadening of 
the HZ is quite modest.  

Remarkable differences exist
between the low-pressure and high-pressure regimes.
At low pressures ($p \lsim 0.3$\,bar) 
the habitability undergoes intense variations in the plane $(a,p)$,
with a general trend of increasing $h$ with increasing  $p$. 
At high pressure ($p \gsim 1$\,bar)  the habitability is approximately
constant and high, with sudden transitions from
$h\simeq 1$ inside the HZ, to $h\simeq 0$. 
%
%
%
The mean planet temperature also shows different characteristics  
between the low  and high pressure regimes.
Starting from pressure $p \gsim 0.3$\,bar, the curves of equal temperature tend to
move  away from the  star as the pressure increases.
This behavior is not seen  at  lower pressures, 
where the HZ  at a given temperature
does not significantly change  its distance from the star. 

Another interesting feature of Fig. \ref{map_Thap_ptot} is the location
of the line of constant mean planet temperature $T_m=273\,$K, indicated
as a magenta solid line superimposed on the symbols of habitability. 
On the basis of the liquid water habitability criterion one would expect
a coincidence of this line with the outer edge of the HZ.
This is true at $p \gsim 2$\,bar, but not at lower values of pressure,
the mismatch being quite large at the lowest values of pressure considered.
The reason is that, using an EBM, we can determine if some latitudinal zones have temperatures larger than zero, even when the mean planet temperature is lower. When this is the case, the planet is partly habitable.

Further insight on the differences between low and high pressures
is offered by Fig. \ref{verticalCrossCut}, where we plot
the surface habitability and temperature as a function of $p$. 
The three panels of the figure show the results obtained at three constant
values of semi-major axis (i.e. insolation). 
In addition to the habitability $h$ (lower solid curve) 
and the mean planet temperature $T_m$ (upper solid curve),
we  show the  minimum and maximum planet temperature
 dot-dashed and dashed curves, respectively).
These temperatures are measured regardless of the time of the year or of the latitude, 
i.e. the maximum temperature could be found at low latitudes and in summer, 
while the minimum, at high latitudes in winter.
The general rise of mean temperature and habitability with increasing
pressure is clear in each case.
Moreover, one can see that 
the excursion between  extreme temperatures 
is quite large at low pressure ($\Delta T \sim 100 - 200$\,K),
but becomes increasingly smaller as the pressure rises. 
Planets with high atmospheric pressure  have a rather uniform
surface temperature. 
%
%

The rich phenomenology described above results from the interaction
of distinct physical factors that we now discuss. 


\subsection{Interpretation} 

The links between surface pressure and habitability are rather intricate.
Pressure variations influence both the planet temperature and 
the extent of the liquid water temperature interval
used to define the habitability. 
To disentangle these effects we first discuss 
the influence of pressure on the temperature and then on the habitability.


\subsubsection{Surface pressure and planet temperature}

Variations of surface pressure  affect the temperature in two ways.
First, for a given atmospheric composition, 
the infrared optical depth of the atmosphere 
will increase
 with pressure. 
As a result, 
a rise of $p$ will always
lead to a rise of the greenhouse effect and    temperature.
Second, the horizontal heat transport increases with pressure.
In our model, this is reflected by the linear increase with $p$ of the diffusion coefficient
$D$  [Eq. (\ref{ModulatedDiffusion1}), Section \ref{sect_D}].
At variance with the first effect, it is not straightforward to predict  
how the temperature will react to a variation of the horizontal transport. 
 
 %
In the case of the Earth, 
our EBM calculations predict a rise of the mean temperature with increasing $D$.
This is due to the fact that 
the increased diffusion from the equator to the
poles tends to reduce the polar ice covers and, as a consequence, to reduce the albedo and raise the temperature.
However,  our calculations predict the existence of particular types
of  climates in which a higher diffusion  yields a lower mean temperature.
An example is shown
in the left panel of Fig. \ref{specificExamples},
where we show the mean annual temperature-latitude profile
for a planet 
at $a=1.0$\,AU with $p =10^{-1.2}$\,bar.
%
Most of the planet is frozen, with the 
exception of an equatorial belt.
In these conditions, an increase in the  diffusion will decrease the equatorial temperature
and extend the ice cover toward low latitudes;
in turn, the increase of the ice cover will cool the planet via  albedo feedback. 
Detailed EBM calculations of this type     
indicate that the initial ice coverage plays a key role
in determining the response of $T_m$  
to a variation of the latitudinal transport.
When the initial ice cover is modest, an increase of $D$ heats the planet, as in the case of the Earth.
When the initial ice cover is somewhat above $\approx 50\%$, an increase of $D$ might cool the planet, 
as in the example discussed above; in these cases the cooling may lead to
a snow-ball transition driven by  the ice-albedo feedback.
 

The total effect of pressure variations on temperature 
will depend on the relative strengths of the greenhouse and latitudinal diffusion effects
discussed above. 
If the greenhouse effect dominates, an increase of $p$ will always
raise the temperature at a given $a$, i.e. at a given insolation
$q = L_\star / (4 \pi a^2)$. 
The results that we find in Fig. \ref{map_Thap_ptot}  
are in line with this expectation
when the pressure is sufficiently high, specifically at $p \gsim 10^{-0.5}$\,bar. 
However, when the atmospheric column density of greenhouse gases is sufficiently low,
pressure variations will not significantly affect
the optical depth and  temperature. 
This probably explains the absence of a temperature rise with increasing $p$
in the regime of very low pressures, visible in Fig.\,\ref{verticalCrossCut}.
%
In this regime, 
an increase of pressure and diffusion may cool the planet.

%
%
%
%

As shown in Fig. \ref{verticalCrossCut},
pressure variations not only influence the mean temperature, but also
the excursion between minimum and maximum planet temperature.
The fact that this excursion becomes increasingly
smaller with rising $p$ is due to
the increase of diffusion efficiency at high pressures. 
The high efficiency of the horizontal transport tends to cancel
seasonal and latitudinal variations in the surface temperature
of the planet. At the other extreme, planets with low atmospheric pressure 
do not benefit from this heat distribution and undergo large excursions
in surface temperature at different latitudes and orbital phases. 

For completeness, we mention that pressure variations also affect
the thermal inertia of the atmosphere
and, as a result, the time scale of adjustment of the
planet temperature to seasonal and latitudinal variations of the radiative forcing. 
In our model this effect is incorporated in Eq. (\ref{Catm})
discussed in Section \ref{sectThermalCapacity} of the appendix. 
This effect becomes important
only when the fraction of oceans is modest.

\subsubsection{Surface pressure and planetary habitability}

%
The location of the inner and outer edges of the HZ is related
to the  boiling and freezing points of water, respectively. 
The temperature of the
 boiling point increases with $p$, while that of the freezing point is basically constant.
As the pressure increases, neglecting any other factor, 
one would expect the inner edge of the HZ
to approach the star  and the outer edge to stay at $a \approx$ constant.
The  inner edge of the HZ  in Fig. \ref{map_Thap_ptot}
confirms this expectation.
The outer edge, instead, moves away from the star as the pressure increases.
This is due to the pressure-temperature effects
described above: at high pressure the greenhouse effect becomes more important
with increasing $p$
and the planet  can remain above the freezing point 
at increasing $a$.
At the inner edge the rise
of the boiling point dominates  over the pressure-temperature effects. 
For considerations about the kind of life that can be expected at these
high temperatures, see Section \ref{whichlife}.

At low pressures the situation is quite complicated.
When $p \lsim 2$\,bar, the contour with $T_m=273\,$K  (magenta line in the figure)
does not overlap with the outer edge of the HZ. 
In fact, there is an area of the plane $(a,p)$ where planets are habitable 
even if $T_m$ is below  freezing point. 
This  can be understood
from the analysis of the latitude-temperature profiles
of planets lying in this area of the map. 
The example at $a=1.0$\,AU with $p =10^{-1.2}$\,bar 
(left panel of Fig. \ref{specificExamples}) explains
 this apparent discrepancy.
As one can see,
even if the mean temperature $T_m$ (solid horizontal line)
is below  freezing point (lower horizontal dashed line)
the existence of a tropical zone of the planet with temperatures above 
 freezing point yields a  habitability $h=0.33$. 

Another peculiar feature of Fig. \ref{map_Thap_ptot}
is  the existence of planets with $T_m$ well inside the
liquid water range but with low levels of habitability.
An example of this type is shown
in the right panel  of Fig. \ref{specificExamples}.
The mean temperature $T_m$ (solid horizontal line)
lies between the freezing and boiling point (horizontal dashed lines), but
the habitability is  only $h = 0.24$.  
The temperature-latitude profile explains the reason for this low habitability: 
most of the planet surface lies outside the liquid-water range
because the equatorial belt is above the boiling point and the high latitude zones
below the freezing point.

These examples clearly indicate that the mean planet temperature 
is not a good indicator of habitability when the planet pressure is low.

%

The pressure dependence  of the horizontal heat transport 
also plays an important role in determining the 
characteristics of habitability. As discussed above, 
planets with high pressure  have a uniform surface temperature
as a result of the high diffusion. The consequence in terms of habitability
is that all the planet surface is either within or outside the 
liquid water temperature range.  If a planet with high pressure
lies inside the HZ, its habitability will be $h \simeq 1$. 
If the temperature goes outside the liquid water range, 
all the planet surface will become un-habitable. 
This explains the sudden transitions from $h\simeq 1$
to $h \simeq 0$ that we see in the upper part of Fig. \ref{map_Thap_ptot}
when we go out of the HZ. 

\subsection{Effects of physical quantities  constrained by observations}

In addition to the semi-major axis,  the orbital eccentricity and the stellar properties can be
measured in the framework of observational studies 
and are relevant for the climate and habitability of exoplanets. 
The results shown in  Fig. \ref{map_Thap_ptot} have been derived for circular orbits and  solar-type stars.  
Here we discuss the extent to which we can generalize these results  
for eccentric orbits and  non-solar type stars.

\subsubsection{Eccentric orbits \label{subsect_ecc}}

To investigate the dependence of the results on orbital eccentricity 
we produced a map of habitability in the plane ($a$,$e$) for
an Earth-like planet with surface pressure $p=1$\,bar. 
For the ocean fraction we adopted $f_o=0.7$ constant at all latitudes.
All the remaining parameters were kept constant, with the values
indicated in Table \ref{multirunPar}.
A total number of 2544 simulations were run
to cover the interval of semi-major axis $0.6 \leq a \leq 1.56$
with a step $\delta a = 0.02$ 
and the interval of eccentricities $0 \leq e \leq 0.95$ with a step $\delta e = 0.02$. 

%

In  Fig. \ref{maps_ae}, we show the resulting   map of habitability and  
mean planet temperature 
in the plane  $(a,e)$.
A characteristic feature of this map is that the 
distance of the HZ increases with increasing eccentricity. 
To interpret this effect we recall that
the mean annual flux received by a planet in an elliptical orbit with semi-major axis $a$
and eccentricity $e$ varies according lo the law (see e.g. Williams \& Pollard 2002)
\begin{equation}
\leftmean q \rightmean = { L_\star \over 4 \pi a^2 (1-e^2)^{1/2}} ~.
\label{meanEccentricFlux}
\end{equation}
Therefore, compared to a circular orbit of radius $a$
and constant insolation  $q_0 = L_\star / 4 \pi a^2$, 
the mean annual flux in an eccentric orbit increases with $e$ according to
the relation $\leftmean q \rightmean = q_0 (1-e^2)^{-1/2}$.
In turn, this increase of mean flux 
is expected to  raise the mean temperature $T_m$.
This is indeed what we find in the results of the simulations. 
The rise of $T_m$   at constant $a$ can be appreciated in the figure, where
the symbols are color-coded according  to $T_m$.
To test this effect in a quantitative way, we superimpose on the figure
the  curve of constant mean flux $\leftmean q \rightmean=q_0$,
calculated from Eq. (\ref{meanEccentricFlux}) for $L_\star=L_\odot$.
One can see that  the HZ 
follows the same type of functional dependence, $e \propto (1-a^{-4})^{1/2}$,
of the curve calculated at constant flux.   This result confirms that the  
increase of mean annual flux is the main effect that governs the shift of the HZ
to larger distances from the star as the eccentricity increases.  

A second characteristic feature of   Fig. \ref{maps_ae}
is that the habitability tends to decrease with increasing eccentricity.
This effect is more evident when we consider the map of 
continuous habitability, $h_c$, in the right panel of the figure.
The effect is related to the large excursion of the instantaneous stellar flux
along  orbits that are very elongated.
The maximum excursion of the flux grows as $\left[(1+e)/(1-e)\right]^2$,
and therefore exceeds one order of magnitude when $e > 0.5$. 
As a consequence of this strong flux variation,  
the fraction of orbital period  in which the planet is habitable at a given latitude
must become increasingly smaller as the orbit becomes more elongated. 
This is equivalent to saying that $f_\mathrm{time}(\varphi)$ decreases
and therefore also $h$ and $h_c$  decrease  (Section \ref{sectHabIndex})
with increasing $e$. 
The effect on $h_c$ must be stronger because this quantity
 depends on $f_\mathrm{time}(\varphi)$ via Eq. (\ref{defHabconTwo}).
The comparison between the left and right panels of Fig.\,\ref{maps_ae}
indicates the existence of an area of the plane $(a,e)$, at high values of $a$  and $e$, populated
by planets that are habitable in small fractions of their orbit.
This is demonstrated  by the fact that such a population disappears when we consider 
the continuous habitability $h_c$.
Apart from 
the existence of this area, it is clear from these figures
that 
the radial extent of the HZ tends to decrease with increasing eccentricity.

On the basis of the above results, it is clear that
if the eccentricity is relatively small, the 
radial extent of the HZ can be scaled according to the mean annual flux.
To this end, we introduce  the {\em effective circular semi-major axis}
\begin{equation}
a_\text{\em eff} = a (1-e^2)^{1/4} ~~~.
\label{effectiveRadius}
\end{equation}
Planets in eccentric orbits have a mean annual flux 
$\leftmean q \rightmean = { L_\star / 4 \pi a_\text{\em eff}^2}$.
%
%
In Fig. \ref{aeff_hab}, we plot several curves of habitability 
calculated at different values of eccentricity
using $a_\text{\em eff}$ in the abscissa.
One can see that, as long as the eccentricity is small ($e \lsim 0.5$)
the curves of habitability versus $a_\text{\em eff}$
calculated at different eccentricities show a good overlap. 
At high eccentricity  the curves of habitability do not overlap. 
The strong effect of eccentricity on the continuous habitability $h_c$ 
is  evident in the right panel of Fig.\,\ref{aeff_hab}. 
However, even using the index $h_c$, the curves of habitability versus $a_\text{\em eff}$
are almost independent of $e$ when the orbits have low eccentricity.

We conclude that, in studies of planetary habitability, 
it is possible to use $a_\text{\em eff}$ as a proxy of $a$  
as long as  the eccentricity is sufficiently small. 
In this way,
we can explore the effects on habitability of physical quantities other than $a$ and $e$
while varying a single parameter that
conveys the information of the orbital characteristics more relevant to the habitability.
However, for highly elongated orbits with $e>0.5$,   one should perform a specific calculation
taking into account both $a$ and $e$. 

The influence of orbital eccentricity 
on planet temperature and habitability has been previously
investigated by Williams \& Pollard (2002) and Dressing et al. (2010).
For comparison with these studies, we  calculated
the mean global planet temperature as a function of orbital longitude
for orbits with $e=$0.1, 0.3 and 0.4.
As a result, we find seasonal trends
similar to those presented in Fig. 2 of Williams \& Pollard (2002)
and Fig. 3 of Dressing et al. (2010); temperature differences are present, but generally lower than 5\,K.  

For comparison with the habitability study of Dressing et al. (2010),
we calculated the temporal habitability, $f_\text{time}$, versus latitude
in orbits of various eccentricities and increasing semi-major axis.
In this way we used our model to reproduce the left panel of Fig.\,7 of Dressing et al. (2010).
In common with these authors  we find that:
(i) the  range of habitable latitudes  shrinks as $a$ increases
and
(ii) the temporal habitability undergoes a transition from high to low values above a critical value of $a$
(the ``outer edge'').
At variance with 
Dressing et al., we find  
(i) a slightly larger value of the outer edge 
(e.g. $1.17\,$AU instead of $1.12\,$AU for the case $e=0.6$)
and 
(ii) a tail of  low temporal habitability ($f_\text{time} \lsim 0.2$) outside the ``edge''.

\subsubsection{Non-solar type stars \label{sectNonSolarStars}}

The luminosity, $L_\star$, and mass, $M_\star$, of the central star 
are free parameters of our climate model.
In the simulations presented in this work, we have
adopted the solar values $L_\star=L_\odot$ and $M_\star=M_\odot$.
In principle, these values can be changed
to calculate  the habitability
of planets orbiting  stars different from the Sun.
%
%
%
%
%
As far as the luminosity is concerned,
it is not even necessary to repeat the calculations: one can 
scale the maps of habitability according to
the level of insolation $q=L_\star/(4 \pi a^2)$, as shown 
in top horizontal axis of Fig. \ref{map_Thap_ptot}. 
In addition, one could incorporate in a single parameter, 
$\leftmean q \rightmean=L_\star/(4 \pi a_\text{\em eff}^2)$,
all the observable quantities that determine the insolation,
including the eccentricity.  
In practice, however, a change of luminosity  implies a change of stellar mass and spectral type
and therefore one should take into account the following effects before applying
the present results to non-solar stars.

For a given value of semi-major axis, 
$a$, the orbital periods of planets in Keplerian orbits around 
stars with different masses 
will scale as $P \propto M_{\star}^{-1/2}$.
As a consequence, for a given level of insolation, the climate simulation should be 
calculated at the proper value of $M_\star$ in order to use the correct orbital period.
Only in this way it is possible to follow the temporal evolution 
of the planet temperature, properly taking   into account the  different time
scales  relevant to the climate system. 
Specifically, the orbital period determines the seasonal evolution of
the incoming stellar radiation, while the relative proportion of oceans and lands
govern the thermal inertia of the climate. 
For habitable planets around stars more massive than the Sun, 
the orbital period is larger than the highest time scale of thermal inertia
(that of the oceans).  The opposite is true for habitable planets around stars with mass
equal or lower than the solar mass.   
 


The extension of the EBM model to low mass stars faces
the problem of the tidal locking. 
The physical quantities in the zonal EBM are meant to be
diurnal zonal averages.
The diurnal average 
is   appropriate  
if the rotation period is  smaller than the orbital period
($P_\mathrm{rot} \ll P_\mathrm{orb}$)
and 
therefore should not be employed when the planet is tidally locked
to its central star ($P_\mathrm{rot} = P_\mathrm{orb}$).
This fact  limits the study of the habitability around
the coolest stars (of M type), since in this case the habitable
zone is generally so close to the star that the planet is expected to become tidally locked
in the long term. 
This limitation does not affect the results that we have presented here,
which refer to the habitability around stars with solar-type  flux. 
The method that we use can be extended to stars cooler than
the Sun, as long as the region of habitability is more distant  than the   
tidal lock radius. According to previous
calculations, the HZ lies beyond the tidal lock radius 
up to spectral type around mid-K  (see e.g. Fig. 5 in Kasting \& Catling, 2003).

In addition to the mass and luminosity, 
the  
spectral type of the central star 
may affect our analysis
in different ways, mainly because the diffusion equation (\ref{diffusionEq})  lacks   
an explicit treatment of the wavelength dependence
of the physical quantities. 

The spectral distribution 
of the  stellar radiation can affect the albedo properties (Selsis et al. 2007).
The prescriptions that we adopt to model the albedo
are calibrated for the Earth and, implicitly,
for a solar-type spectral  distribution. 
As discussed above for the tidal locking problem,
the conclusions that we derive can be  extended to stars 
with a somewhat later spectral type.  
However, care should be taken in applying our results to M-type stars. 

A further issue related to the wavelength dependence  concerns
the separation between the incoming radiation, $S$,
and the outgoing radiation, $I$, in Eq. (\ref{diffusionEq}).
This formal separation
is valid if the spectral distribution of these two terms
is well separated in wavelength. 
This condition is well-satisfied for solar-type stars since the radiation of a G2 star
peaks at $\simeq 0.5\,\mu$m and that of a habitable planet ($T \sim 300$\,K)  
at $\simeq 10\,\mu$m.
For stars of later spectral type, the peak  shifts to longer wavelengths,
but the assumption is still reasonable (the radiation of a M5 star peaks at $\simeq 0.9\,\mu$m).

\subsection{Effects of parameters unconstrained by observations}

At variance with the eccentricity and the stellar luminosity 
that can be derived from observational methods,
many of the planetary parameters relevant for 
the climate and habitability are unconstrained by the observations
of extrasolar planets. 
This is true, for instance, for  
rotation period, axis obliquity, planet geography, and surface albedo. 
Here we discuss how variations
of such unconstrained quantities,
in combination with variations of surface pressure, 
may affect the habitability of exoplanets



\subsubsection{Planet rotation period \label{sectRotationPeriod}}

In Fig. \ref{ahp_prot} we plot the curves of habitability $h$ versus $a$
obtained for three  rotation periods:
$P_\text{rot}=1/3$\,d, 1\,d, and 3\,d 
(dashed, solid and dashed-dotted curves, respectively). 
The four panels of the figure 
correspond to $p=0.1$, 0.3, 1.0, and 3.0\,bar, as indicated in the labels.  

Planet rotation affects the  habitability curves.
The area subtended by the curves tends to increase and the shape tends
to become top-flatted as the rotation period increases. 
The interpretation
of this effect is as follows. 
An increase of $P_\text{rot}$ yields a quadratic increase of the 
diffusion  via Eq. (\ref{ModulatedDiffusion1})
because $D \propto \Omega^{-2} \propto  P_\text{rot}^{2}$. This effect  tends to
homogenize the surface temperature,  particularly in the high pressure regime
(bottom panels), since the diffusion coefficient increases  with $p$ as well. 
An homogeneous temperature will yield abrupt transitions
between a fully habitable and a fully non-habitable situation, the extreme
case considered here being the top-flatted, box-shaped habitability curve
at $P_\text{rot}=3$\,d and $p=3.0$\,bar in the bottom right panel of the figure.
At the other extreme of low rotation periods and low $p$, the habitability curves
show instead a pronounced peak. 
Planets in these conditions experience significant variations in their
habitability even with a modest change of $a$ (i.e., of insolation). 
When the pressure is  low (top-left panel) the habitability is relatively low.
An increase of the rotation period helps to transport the horizontal heat
and  tends to raise the  peak and increase the area 
subtended by the curve. 

The comparison between the results shown in the four panels of Fig. \ref{ahp_prot}
indicates that the width and centroids of the curves of habitability  
tend to change together as pressure varies. 
All together, variations of the rotation period
do not dramatically affect the general features  of the pressure-dependent  HZ.

Previous studies on the influence of the rotation period on planetary climate and habitability
have been presented by SMS08. In common with these studies, we find a  decrease
in the planet temperature and habitability with increasing rotational velocity.
However, our calculations do not yield the runaway transition to a snowball climate that was found by SMS08
by adopting $P_\text{rot}=1/3$\,d for the Earth. 
The different choice of the albedo prescriptions is a plausible reason for this different result.
As one can see in the right panel of
Fig. \ref{earthLatTemp},  the albedo-latitude profile adopted by SMS08 is characterized
by a sharp transition at high latitudes. Our model profile, which
is in better agreement with the experimental data, does not show such a feature. 
The sharp transition of the albedo adopted by SMS08 is probably responsible
for the very strong ice-albedo feedback found by these authors.

\subsubsection{Axis obliquity}

In   Fig. \ref{ahp_obliq}, we plot the curves of habitability $h$ versus $a$
calculated at  4 values of  obliquity:
$\epsilon=0^\circ$, $30^\circ$, $60^\circ$, and $90^\circ$
(dashed, solid, dashed-dotted, and dotted curves, respectively). 
As in the previous figure, the four panels 
correspond to $p=0.1$, 0.3, 1.0, and 3.0\,bar. 

The curves of habitability 
show strong variations with obliquity, in particular when the pressure is low. 
The  general trend, at all pressures, is that the habitability
increases with increasing $\epsilon$ when the obliquity is low
($\epsilon \lsim 60^\circ$)
and decreases with increasing $\epsilon$ when the obliquity is high
($\epsilon \gsim 90^\circ$). 
This complex behavior can be explained in the following way. 
The configuration at $\epsilon =0^\circ$ favors the formation of  permanent ice caps
in the polar regions, where the star is always at large zenith distance.  
As the obliquity starts to increase from zero, 
a larger fraction of polar regions undergo a period of stellar irradiation 
at low zenith distance in some phase of the orbit. This tends to reduce
the ice caps and therefore to increase the habitability. 
However,  when the obliquity becomes quite large, a permanent ice belt starts to build up
in the equatorial zones, leading to a decrease in the habitability. 
To understand why the equator is colder than the poles we consider the extreme case $\epsilon=90^\circ$.
%
In this case, the maximum
insolation of a polar region occurs when the pole faces the star;
the instantaneous insolation at the pole does not change
during the planet rotation, so that the mean diurnal insolation is $S=q$.
The maximum insolation of an equatorial region instead occurs when the rotation axis is
perpendicular to the star-planet direction.
In this case the instantaneous insolation at the equator undergoes
the night-day cycle and the mean diurnal flux is  $S=q/\pi$
[see Eq. (\ref{meanDiurnalFlux}) for this specific configuration, 
in which $\delta=0$,   $\varphi=0$, and $H=\pi/2$]. 
Our calculations indicate that the 
equatorial zones start to build up permanent ice
when the obliquity increases from $\epsilon=60^\circ$ to $\epsilon=90^\circ$.
The exact seasonal evolution is strongly dependent on the thermal inertia of the climate components.

The formation of an equatorial ice belt was first predicted by WK97 for the case
$\epsilon=90^\circ$ and $p=1$\,bar. 
Our calculations indicate that the effect is  stronger at lower pressures. 
%
%
The obliquity effects tend to disappear as the pressure increases.
The bottom panels of  Fig. \ref{ahp_obliq} show that
at $p=1$\,bar, the variations are moderate  
and at $p=3$\,bar, the influence of obliquity becomes modest.
This is due to the high efficiency of the horizontal transport
at high $p$, which tends to cancel temperature gradients on the planet surface,
preventing the formation of polar ice caps or an equatorial ice belt
at extreme values of obliquity.

The width and centroids of the curves of habitability  
calculated at different obliquities
tend to change together as pressure varies
in the four panels of Fig. \ref{ahp_obliq}. 
In this respect, we can conclude,
as in the case of the rotation period, 
that variations of the obliquity do not affect the overall characteristics of 
the pressure-dependent HZ. 

Previous studies on the effects of obliquity have been performed by WK97
and Spiegel et al. (2009; hereafter SMS09). 
For comparison with WK97, we  
calculated the zonal surface temperature, $T$, as a function
of orbital longitude, $L_\text{S}$, for
an Earth with obliquity $\epsilon=90^\circ$.
By considering the temperatures in five latitude zones 
centered on latitudes $-85^\circ$, $-45^\circ$, $+5^\circ$, $+45^\circ$, and $+85^\circ$,
we obtain trends of zonal $T$ versus $L_\text{S}$
very similar to those shown in Fig.\,1B of WK97, with temperature differences  
generally below 10\,K. 

For comparison with SMS09, we calculated the 
temperature-latitude profile of a planet with the same North Polar continent
considered by these authors, for three different obliquities
($\epsilon=23.5^\circ$, $60^\circ$ and $90^\circ$).
Our model predicts 
temperature-latitude profiles similar to those visible
in Fig.\,8 of SMS09, 
with an important difference:  in the case $\epsilon=90^\circ$
the temporal evolution that we find is similar to that 
found by SMS09 only in the first $\simeq 10$ orbits of the simulation.
Afterwards, a snowball transition starts to occur in the simulation of SMS09
(bottom-right panel of their Fig.\,8),
but not in our simulation. 
This result confirms that our climate model is relatively stable againts 
snowball transitions, 
as we have
discussed at the end of the previous 
Section \ref{sectRotationPeriod}. 
Apart from this fact, the influence of obliquity on 
 climate predicted by our EBM is similar to that predicted by previous work. 

%
%

\subsubsection{Planet geography \label{sectGEO}}

Climate EBMs incorporate planet geography in a schematic way,
the main parameter being the zonal coverage of oceans, $f_o$,
which  at the same time determines the coverage of lands, $f_l=1-f_o$. 
To explore the  effects of geography on habitability we performed two types of tests. 
First we compared planets with different global coverage of oceans
keeping $f_o$ constant  in all latitude zones.
Then we compared planets with a different location of  continents (i.e. polar versus equatorial)
keeping  constant the global fraction of oceans.

The results of the first test are shown in   Fig. \ref{ahp_ocean}, where 
we plot the curves of habitability $h$ versus $a$
calculated at  3 values of  ocean fraction:
$f_o=0.25$, 0.50, and 0.75 
(solid, dotted and dashed-dotted curves, respectively). 
Each panel shows the results obtained at a constant pressure 
$p=0.1$, 0.3, 1.0, and 3.0\,bar. 
Two effects are visible as the coverage of oceans increases:
(i) the HZ tends to shift to larger distances from the star;
(ii) at low pressures ($p \lsim 1$\,bar)  the habitability  tends to increase.
Both effects are relatively small, at least for the combination of parameters considered.
The increase in the thermal inertia of the climate system 
with increasing global ocean fraction is the key to interpreting these results. 
At low pressure the high thermal inertia of the oceans tends to compensate
the inefficient surface distribution of the heat typical of low-pressure atmospheres.
At high pressure, the combination of a high thermal inertia 
with an efficient atmospheric diffusion tends to give the same temperature
at a somewhat smaller level of insolation. 

To investigate the effects of continental/ocean distribution, 
we considered the three model geographies proposed by WK97: 
(1) present-day Earth geography, (2) equatorial continent, and (3) polar continent. 
In practice, each model is specified by a set of ocean fractions, $f_o$, of each latitude zone
(Table III in WK97). The case (2) represents a continent located
at latitudes $|\varphi| < 20^\circ$ covering the full planet. The case (3) a polar continent
at $\varphi \lsim -30^\circ$. The global ocean coverage is approximately the same 
($\leftmean f_o \rightmean \simeq 0.7$)
in the three cases. 
As a result, we find that these different types of model geographies introduce modest effects on
the habitability curves.  The habitability of present-day and 
equatorial continent geography are essentially identical at all pressures.
The polar continent geography is slightly less habitable.
This is probably due to the combination of two factors that
tend to form a larger ice cap in presence of a polar continent:
(i) ice on land has a higher albedo than ice on water and
(ii) the thermal capacity of continents is lower than that of oceans.
In any case, the  differences in habitability are small and tend to disappear at $p \gsim 3$\,bar  
due to the fast rise of the horizontal heat transport.


\subsubsection{Albedo of the continents \label{sectAlbedoLands}}

Albedo variations can shift the location of the HZ,
moving  the HZ inward if the albedo increases, or outward if the albedo decreases.
In our model,
the albedo of the oceans, ice and clouds are not free parameters
since they are specified by well-defined prescriptions
(Section \ref{sectionAlbedo}). The albedo of the lands, $a_l$, is instead a free parameter. 
In the simulations run to build
the map of Fig. \ref{map_Thap_ptot}  we kept a fixed value $a_l=0.2$, 
representative of the average of Earth continents. In a generic planet,
the albedo of lands can vary approximately between $\simeq 0.1$ and $0.35$,
depending on the type of surface. The lowest values are appropriate, for instance, 
for basaltic rocks or conifer forests, while the highest values for Sahara-like deserts
or limestone; 
Mars sand has $\simeq 0.15$ while grasslands $\simeq 0.2$ (Pierrehumbert 2010).
Given this possible range of continental albedos,
we have repeated our calculations for $a_l=0.1$, $0.2$ and $0.35$.
As a result, we find that the curve of habitability $h$ versus $a$ shift closer to the star
for  $a_l=0.35$ and away from the star for $a_l=0.1$.
The maximum shift between the extreme cases is $\simeq 0.03$\,AU.
The shape of the habitability curves is virtually unaffected by these changes. 

 
\subsubsection{Surface gradient of latitudinal heat transport}

With our formulation of the diffusion coefficient, we can
investigate how planetary habitability is influenced 
by variations of the heat transport efficiency.
In practice, this can be done by varying the  parameter $\mathcal{R}$,
that represents the ratio between the maximum and minimum value of the diffusion
coefficient in the planet (Section \ref{sectDiffCoefficient}). 
The results shown in Fig. \ref{map_Thap_ptot} have been obtained
for $\mathcal{R}=6$, a value optimized to match Earth experimental data. 
We repeated our calculations for $\mathcal{R}=3$ and $\mathcal{R}=12$,
keeping constant the fiducial value $D_0$, that represents
the mean global efficiency of heat transport. 
As a result, we do not find any significant difference in planetary habitability
either at high  or  low pressure. 
We conclude that the knowledge of the
exact functional dependence on the latitude of the heat transport efficiency
is not fundamental in predicting  the properties of the HZ,
at least with the simplified formalism adopted here
that does not consider the circulation due to atmospheric cells.

\subsubsection{ Summary}

The main effects of varying planetary parameters can be summarized as follows. 
As far as the {\em shape} of the habitability curves is concerned,
the results at low pressure are quite sensitive 
to variations of rotation period, axis obliquity and ocean coverage.
Specifically, the habitability tends to increase 
with increasing rotation period, axis obliquity (up to $\epsilon \simeq 60^\circ$), and ocean coverage.  
At high pressure, the shape of the habitability curves become insensitive to 
variations of these parameters.
As far as the {\em location and radial extension} of the habitability curves is concerned,
the results  are modestly influenced by variations of planetary parameters, even at low pressure.
Variations of the latitudinal/seasonal gradient of the heat transport efficiency
do not affect the properties of the HZ.
Albedo variations tend to shift the habitability curves
without affecting their shape.

\subsection{Final considerations}

%

\subsubsection{The edges of the habitable zone \label{CO2cycle}}
 
%

Our calculations do not include the CO$_2$ climate stabilization mechanism
considered in the classic HZ definition (Kasting et al., 1993). 
In principle, EBM models can be used to
simulate the carbonate-silicate cycle that drives this mechanism 
(WK97). 
However, this choice is not practical when running thousands of simulations, 
as we do here,
because the time-scale of the CO$_2$ cycle 
is much larger than the time-scale of convergence of the simulations.
%
In addition,
modeling the carbonate-silicate cycle 
requires 
{\em ad hoc} assumptions on the 
silicate weathering law and rate of CO$_2$ production by volcanos.
At the present time 
it is not clear if an active volcanism and tectonics,
the ingredients required for the existence of the CO$_2$ 
cycle, are common in terrestrial planets.
For planets 
that do have this cycle, 
the outer edge of the HZ would shift to  larger values
of semi-major axis than estimated here. 
For completeness we note that, if oceans are salty, 
the outer edge would shift in the same direction due to the lowering 
of the water freezing point.




As far as the inner edge is concerned,  we adopt
the water-loss limit criterion (\ref{eqTmax}). 
At $p=1$\,bar, our inner edge is located $a=0.82$\,AU,
 in the range of inner limits predicted by Selsis et al. (2007). 
%
However,  the study of the latitude-temperature profiles
obtained from our simulations show the existence of cases 
that challenge the definition of the inner edge.
Specifically,
at low values of $a$ and $p$ in Fig. \ref{map_Thap_ptot},
 we find  planets that are habitable, but have
temperatures above the boiling point of water in some latitude zones.
An example is shown in the right panel of Fig. \ref{specificExamples}.
The water reservoir of these type of planets 
is  likely to undergo a complete evaporation
in the long term. 
In fact, to avoid this fate, 
the continental distribution and axis obliquity should ``conspire'' to keep
the oceans outside the boiling zone 
during all the planet's life.
This in turn would require (i) a long-term mechanism 
of stabilization of the rotation axis
and (ii) the absence of tectonics drifts that, sooner or later, would   
build up a geography prone to evaporation.
%
%
If the water reservoir is lost after an initial period of habitability, 
these planets  match  the definition of ``Class II'' habitats  
proposed by Lammer et al. (2009):
bodies which  possess Earth-like 
conditions 
in the early stages of evolution, but not  in the long term.
According to Lammer's scheme, Venus is a potential ''Class II'' object that may have lost its
water by evaporation.  
In contrast with  Venus,
the planets that we find 
have a low  boiling point, typical of  low pressure, and
can evaporate at 
a relatively cold global temperature.
By excluding these  ``cold evaporating planets''  
from the map of Fig. \ref{map_Thap_ptot},
the low-pressure inner edge would shift  to larger values of 
semi-major axis
(e.g. from 0.87\,AU to 0.93\,AU at $p=0.1$\,bar).

\subsubsection{Which type of life? \label{whichlife}}

\begin{figure}
\begin{center}
\plotone{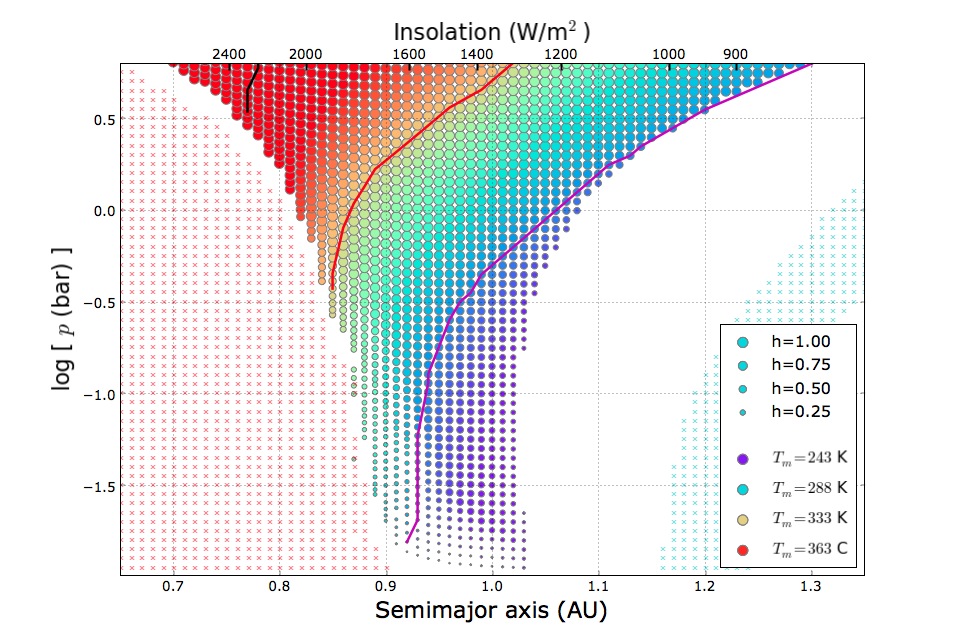} 
\caption{
Circumstellar habitable zone of planets with Earth-like atmospheres
and different levels of surface pressure 
obtained with our EBM climate simulations. Abscissae: semi-major axis, $a$ (bottom axis),
or insolation $q=L_\star/(4 \, \pi \, a^2)$ (top axis). Ordinates: logarithm of the total surface pressure, $p$.
The circles indicate solutions with mean global annual habitability $h>0$.
The area of the  circles is proportional  to $h$; the colors are coded
according to the mean annual global surface temperature, $T_m$.
The  size and color scales are shown in the legend. 
The solid lines are contours of equal mean temperature
$T_m=273\,$K (magenta), $333\,$K (red) and $393\,$K (black).
%
Results above the contour at $T_m=333\,$K (red line) are tentative; see Section \ref{maps}. 
%
%
Red crosses:
simulations stopped on the basis of the water loss limit criterion [Eq.\,(\ref{eqTmax})]; 
blue crosses: simulations interrupted when $T_m < T_\text{min}$;
see Section \ref{running}.
%
Adopted model parameters are listed in Tables \ref{tabCalibratedPar} and \ref{multirunPar}.
\label{map_Thap_ptot}}
\end{center}
\end{figure}

\begin{figure}
\begin{center}
\includegraphics[width=8cm]{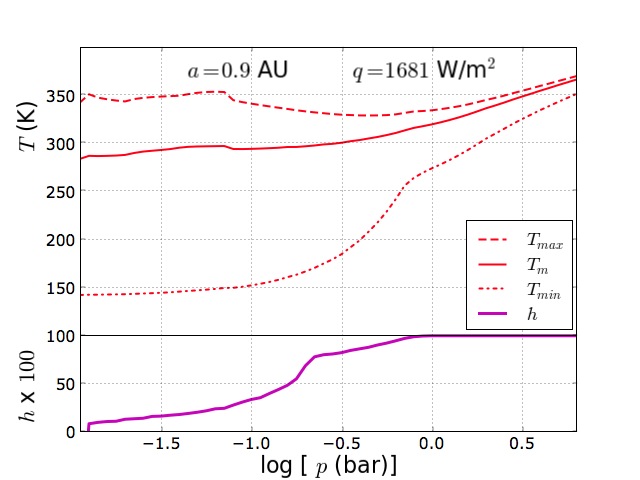} 
\hskip 2cm
\includegraphics[width=8cm]{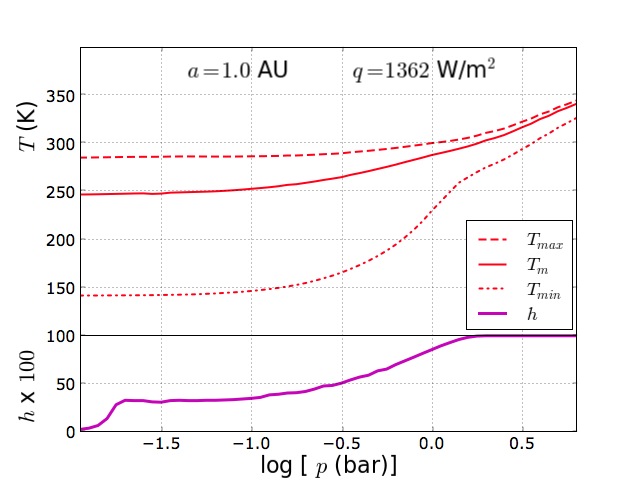} 
\hskip 2 cm
\includegraphics[width=8cm]{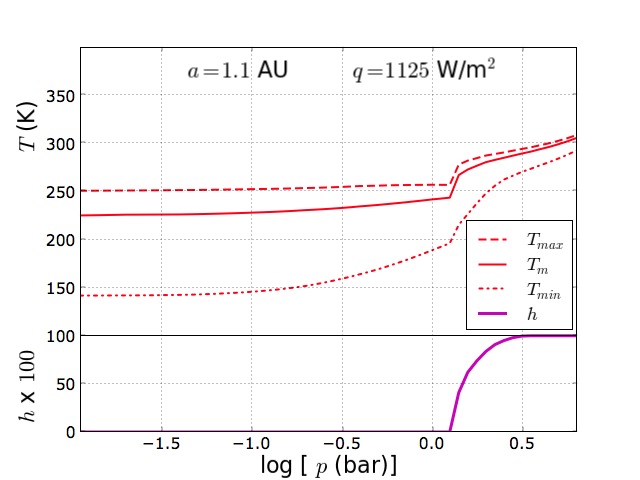}   
\caption{ 
Planet surface temperature, $T$, and habitability, $h$, as a function of 
surface pressure, $p$.
Each panel shows the results obtained
at constant semi-major axis, $a$, and
constant insolation, $q$, indicated in the legend. The solid curve at the bottom
of each panel is the habitability expressed in percent units.
The three curves at the top of each panel are temperature curves in kelvin units
(solid line: mean planet temperature; dot-dashed and dashed lines: minimum and maximum
planet temperatures at any latitude and season). 
Adopted model parameters are listed in Tables \ref{tabCalibratedPar} and \ref{multirunPar}.
\label{verticalCrossCut}}
\end{center}
\end{figure}

\begin{figure}
\begin{center}
\includegraphics[angle=0,width=8cm]{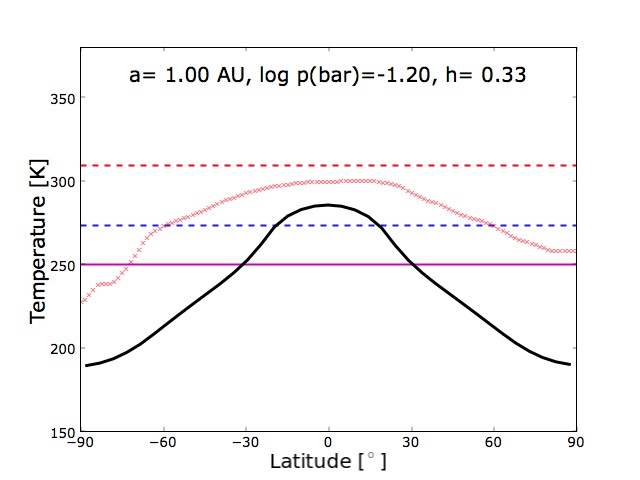}
\includegraphics[angle=0,width=8cm]{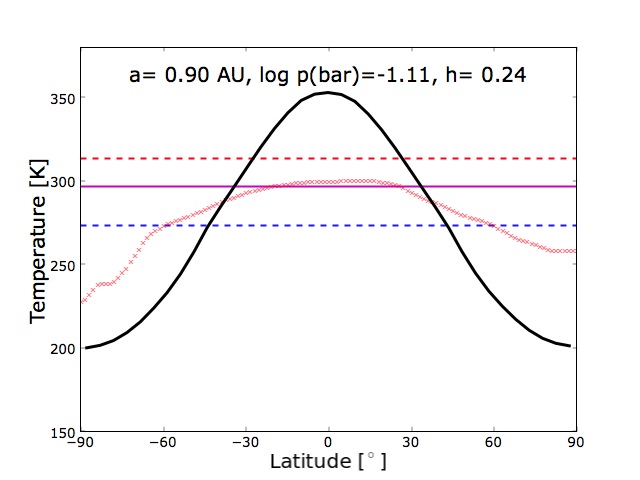}
\caption{Solid curve: 
mean annual temperature-latitude profile of an Earth-like planet
with semi-major axis, $a$,  surface pressure, $p$, 
and habitability, $h$, specified in the legend of each panel.
Solid horizontal line: mean annual global temperature, $T_m$.
Dashed horizontal  lines: liquid water temperature interval at
pressure $p$. 
Adopted model parameters are listed in Tables \ref{tabCalibratedPar} and \ref{multirunPar}. 
Temperature-latitude profiles are symmetric as a result of the idealized
geography used in the simulation (constant fraction of oceans in all latitude zones). 
Crosses:  Earth data as in Fig. \ref{earthLatTemp}.
 \label{specificExamples}}
 \end{center}
\end{figure}

\begin{figure}
\plottwo{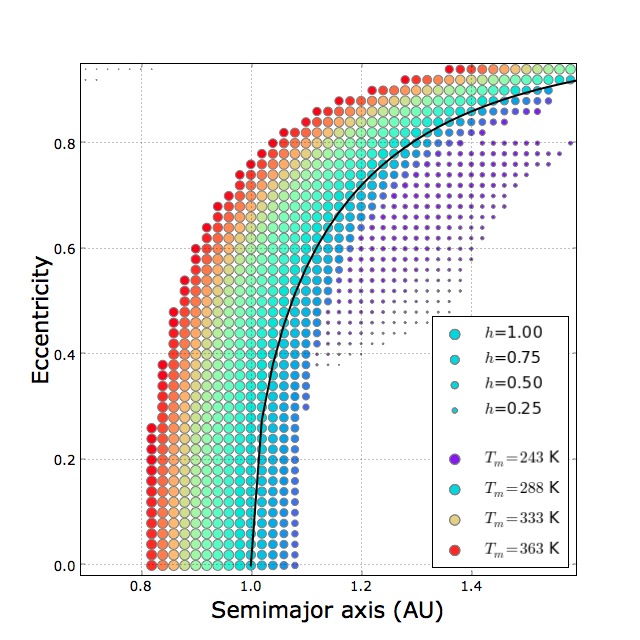}{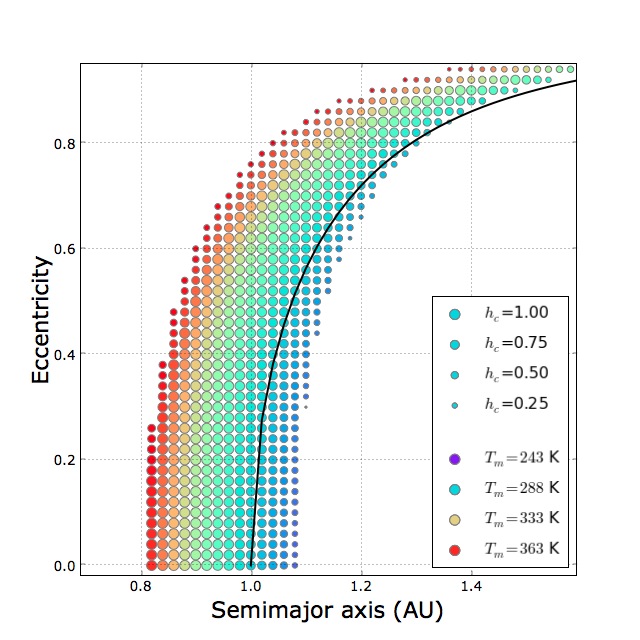}
\caption{Maps of  mean temperature  
and habitability of an Earth-like planet in the plane of the semi-major axis and eccentricity. 
Left panel: habitability $h$;
right panel: continuous  habitability $h_c$ (see Section \ref{sectHabIndex}).
The area of the circles is proportional to the
mean fractional habitability;
the color varies according to the mean annual global surface temperature, $T_m$.
The  size and color scales are shown in the legend. 
%
Adopted model parameters are listed in Tables \ref{tabCalibratedPar} and \ref{multirunPar},
with the exception of the eccentricity
that has been varied as shown in the figure. 
Solid curve: line of equal mean annual flux $\leftmean q \rightmean=q_0$
estimated from Eq. (\ref{meanEccentricFlux}).
\label{maps_ae}}
\end{figure}

\begin{figure}
\plottwo{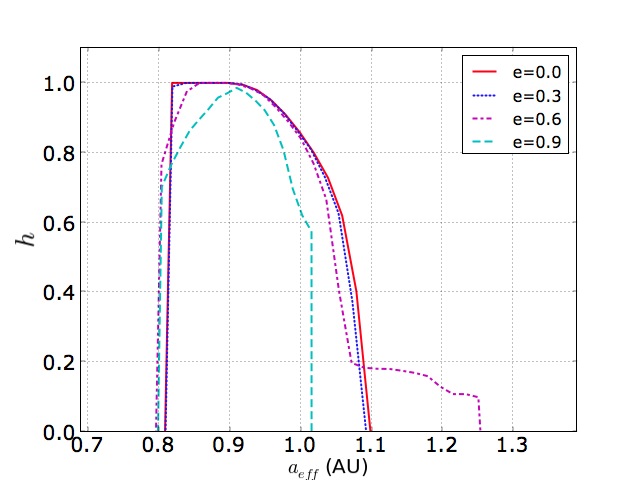}{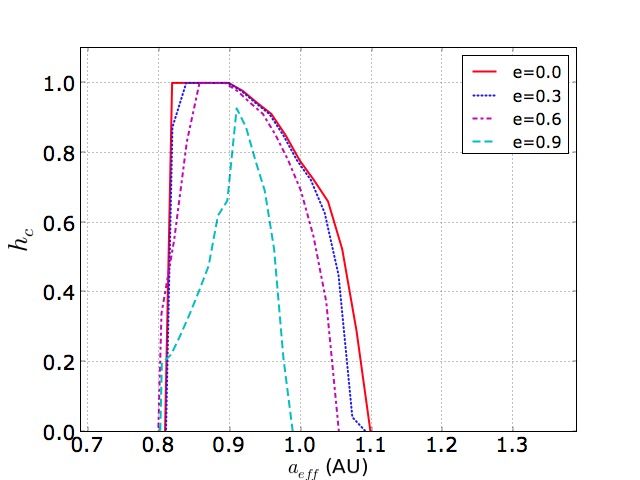}
\caption{ 
Radial extent of the habitable zone for an Earth-like planet
in a keplerian orbit of increasing eccentricity around a star with solar luminosity.
Abscissae: 
effective circular semi-major axis $a_\text{\em eff} = a (1-e^2)^{1/4}$
defined in Section \ref{subsect_ecc}. 
Left panel: habitability $h$;
right panel: continuous  habitability $h_c$ (see Section \ref{sectHabIndex}).
Model parameters as in Fig. \ref{maps_ae}.
\label{aeff_hab}}
\end{figure}

\begin{figure}
\includegraphics[width=8.cm]{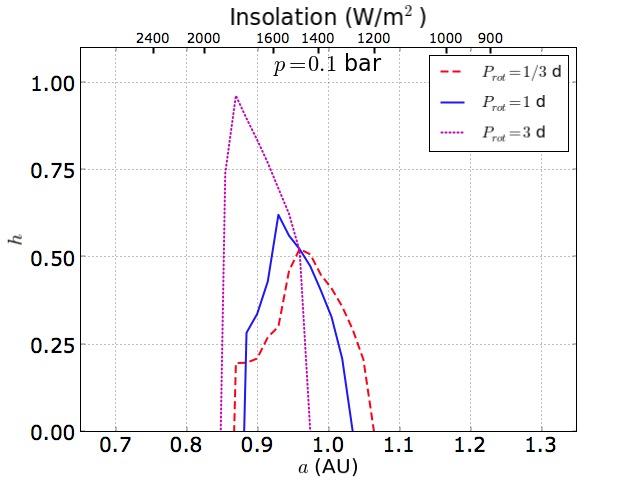} 
\includegraphics[width=8.cm]{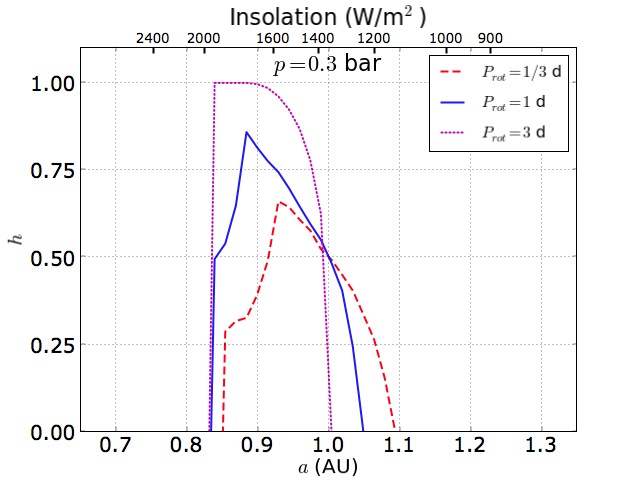} 
\includegraphics[width=8.cm]{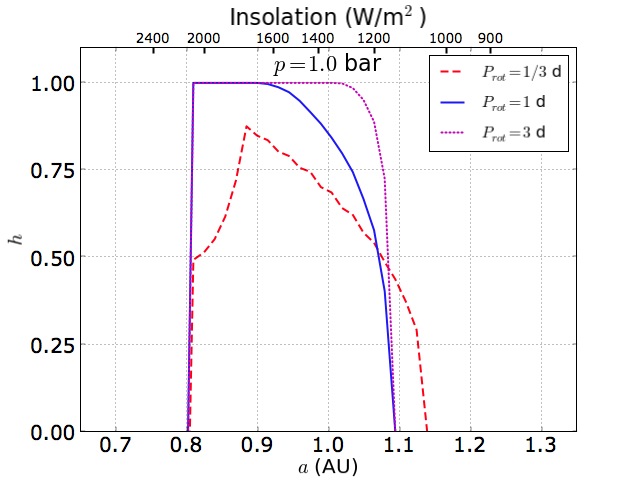}  
\includegraphics[width=8.cm]{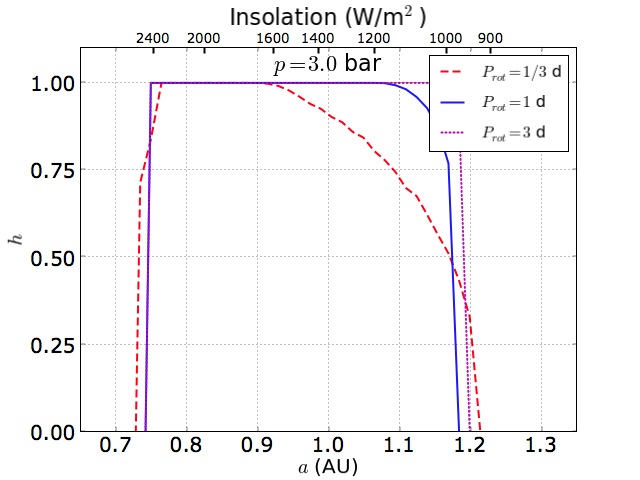} 
\caption{ 
Fractional habitability, $h$, as a function of semi-major axis, $a$,
for planets with rotation periods $P_\text{rot}=1/3$\,d, 1\,d, and 3\,d.
Each panel shows the results obtained at a constant pressure $p$. 
The other parameters of the simulations are listed 
in Tables \ref{tabCalibratedPar} and \ref{multirunPar}.
 \label{ahp_prot}}
\end{figure}

\begin{figure}
\includegraphics[width=8.cm]{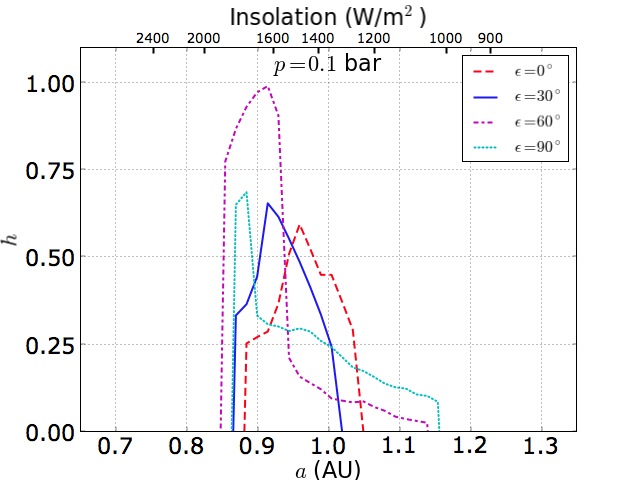}  
\includegraphics[width=8.cm]{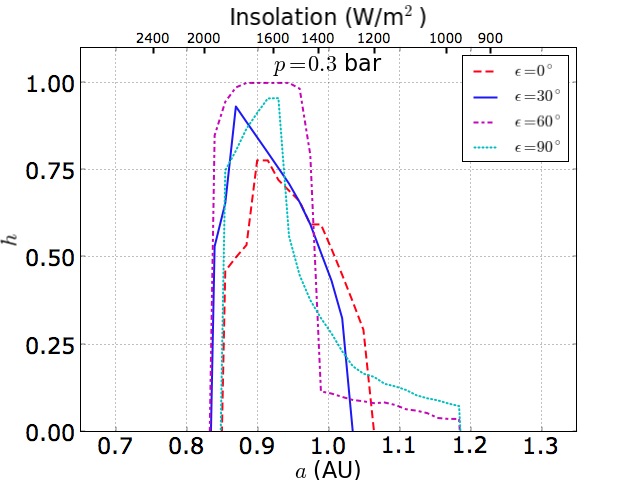}  
\includegraphics[width=8.cm]{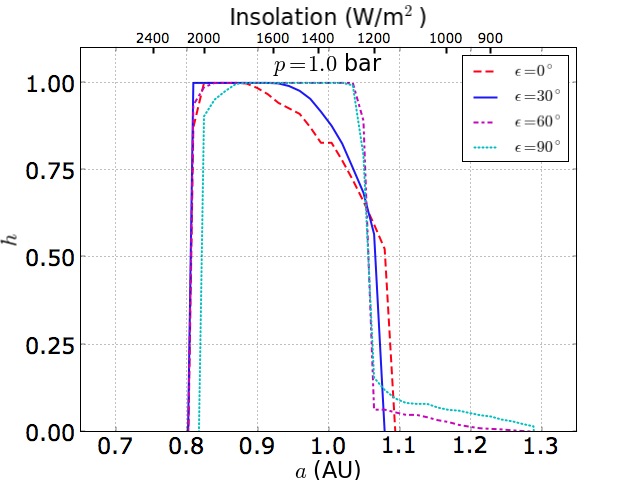}  
\includegraphics[width=8.cm]{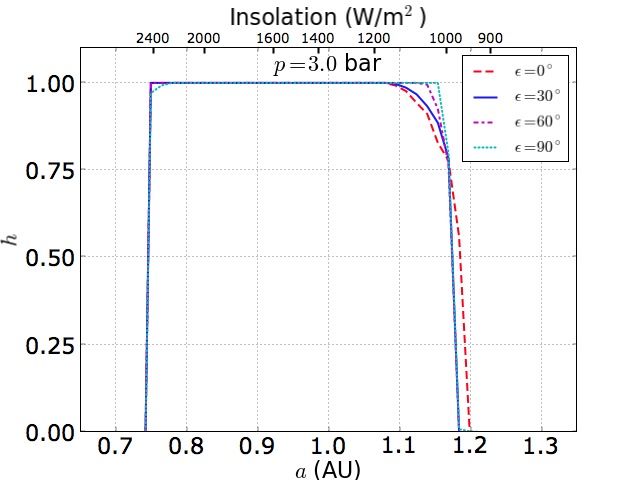}  
\caption{ 
Fractional habitability, $h$, as a function of semi-major axis, $a$,
for planets with axis obliquity $\epsilon=0^\circ$, $30^\circ$, $60^\circ$, and $90^\circ$.
Each panel shows the results obtained at a constant pressure $p$. 
The other  parameters of the simulations are 
listed in Tables \ref{tabCalibratedPar} and \ref{multirunPar}.
 \label{ahp_obliq}}
\end{figure}

\begin{figure}
\includegraphics[width=8.cm]{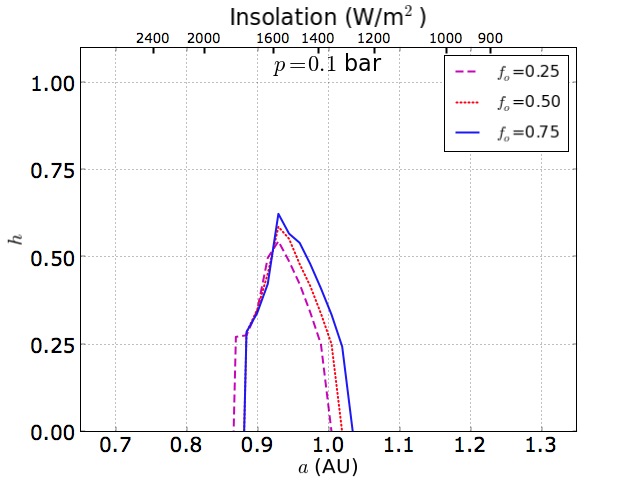}  
\includegraphics[width=8.cm]{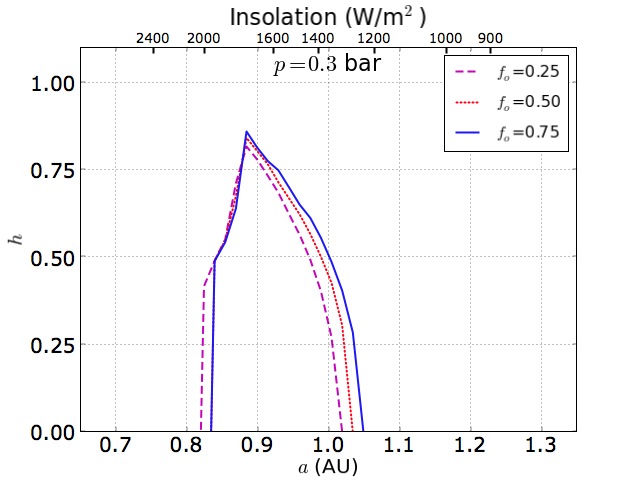}  
\includegraphics[width=8.cm]{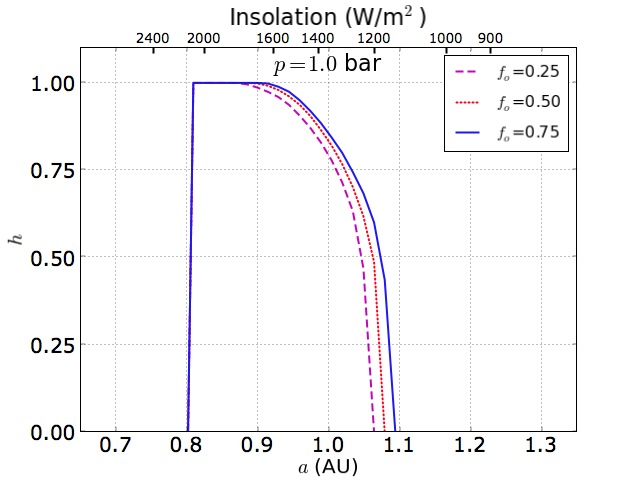}  
\includegraphics[width=8.cm]{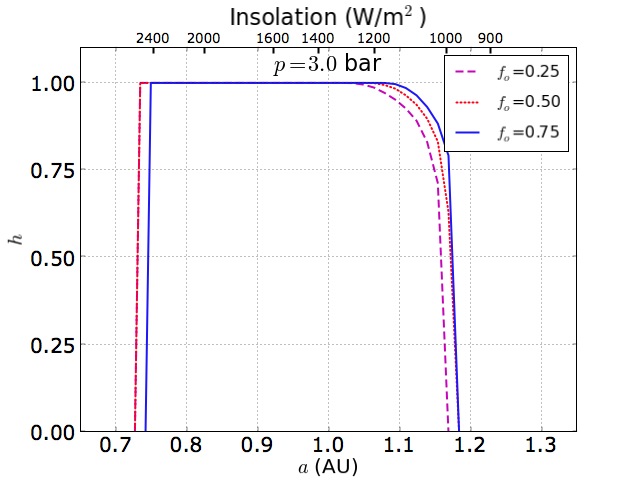}  
\caption{ 
Fractional habitability, $h$, as a function of semi-major axis, $a$,
for planets with ocean fractions $f_o=0.25$, $0.50$, and $0.75$.
Each panel shows the results obtained at a constant pressure $p$. 
The other  parameters of the simulations are 
listed in Tables \ref{tabCalibratedPar} and \ref{multirunPar}.
\label{ahp_ocean}}
\end{figure}

Even if the HZ of Fig. \ref{map_Thap_ptot} broadens with pressure,
the type of life that can be expected to exist at a given level of pressure
$p \gsim 1$\,bar
 may be quite different, depending on the exact location of the planet  in the plane $(a, p$).
Indeed, most of the enlargement of the HZ at high pressures 
is due to the increasing fraction of regions with high
temperatures lying above the isothermal contour $T_m = 60 \celsius$ 
(red curve in Fig. \ref{map_Thap_ptot}).
Taking terrestrial life as a reference, these regions  
would only be habitable by extremophilic organisms,
specifically thermophiles and hyperthermophiles (see e.g. Cavicchioli 2002). 
In fact,
no terrestrial extremophile is known to exist above $\simeq 110-120 \celsius$,
the record being  shared by a few archea living in  oceanic   
hydrothermal vents 
({\em Pyrolobus fumarii}, 
{\em Methanopyrus kandleri strain 116}, and 
{\em Strain 121}). 
%
On these grounds, it is debatable whether forms of life may exist
beyond the  isothermal contour $T_m = 120 \celsius$ 
(black curve in Fig. \ref{map_Thap_ptot}).
If we are interested in the distribution of mesophilic organisms, 
rather than in extremophiles,
the HZ does {\em not} become larger with increasing pressure:
for  organisms adapted to a temperature within the range
$0 \celsius \lsim T_m \lsim 60 \celsius$ the HZ
shifts to larger distances from the central star as the pressure increases
(region between the magenta and red curves in Fig. \ref{map_Thap_ptot}).

As discussed above, the surface temperature becomes quite uniform when
the pressure exceeds a few bars. The uniformity of the temperature
at all latitudes and seasons means that habitable planets with 
high pressure  can only host surface life  adapted to a fine-tuned range of temperatures.
As an example, the maximum temperature excursion 
at $p=5$\,bar at the outer edge of the HZ
is $-12 \celsius \leq T \leq +5 \celsius$. 
This type of situation typical of high pressure is
quite different from that of the Earth, where the latitudinal and seasonal
variations allow the presence of a wide diversity of  surface life 
adapted to different temperatures. 
At lower pressures, the temperature excursions tend
to become quite high allowing, in principle, an even broader   
gamut of surface life.
In particular,  terrestrial-type cryophilic organisms 
with optimal temperature $T < -15 \celsius$ would find 
a proper surface habitat only at low pressure ($p \lsim 1$\,bar)
in the outermost regions of the HZ. 
However, at  very low pressure the overall habitability    
severely decreases and  becomes
restricted  to a  narrow range of distances from the star. 
%
%

%
%

\section{Conclusions}

We have implemented  a 1-D energy balance  model (EBM) of planetary climate,
based on the diffusion equation (\ref{diffusionEq}),
aimed at exploring the surface habitability of extrasolar planets.
Starting from the model prescriptions adopted in previous EBMs, our  model
contains  new recipes for the diffusion coefficient, the outgoing IR flux,
the albedo, the effective thermal capacity, and the ice and cloud cover. 
%
%
Our prescription for the diffusion coefficient  introduces in a natural way 
seasonal and latitudinal variations in the efficiency of the heat transport.
The formalisms adopted for the calculation of the albedo and
effective thermal capacity are sufficiently general to be applied
to planets with a variety of surface conditions. 
%
%
The recipe for the ice cover allows the formation of permanent 
ice caps or belts.
%
The model parameters have been fine-tuned in such a way to reproduce the
mean annual latitude profiles of the Earth temperature and albedo
(Fig.\,\ref{earthLatTemp}), as well as
the seasonal variability of the temperature-latitude profiles
(Fig.\,\ref{earthSeasons}). 

As a first application of our model, we have investigated
the habitability of planets with Earth-like  atmospheric composition
but different levels of surface pressure.
The habitability is  estimated  on the basis of the 
pressure-dependent liquid water temperature range,
taking into account the seasonal and latitudinal variations of the 
surface temperature.
By running a  large number of climate simulations with our EBM 
we have estimated the mean global annual temperature, $T_m$, and habitability, $h$,
as a function of semi-major axis, $a$, and  surface pressure, $p$.
In this way we have built-up a habitability map (Fig. \ref{map_Thap_ptot})
that represents the pressure-dependent HZ for planets with Earth-like atmospheres
orbiting a solar-type star. The main results that we find can be summarized as follows.

\begin{itemize}

\item
The  radial extent of the  HZ increases with pressure,
from $\Delta a=0.18$\,AU at $p=1/3$\,bar,  
to $\Delta a=0.43$\,AU at $p=3$\,bar.

\item
At a given value of semi-major axis $a$, or insolation, $q=L_\star/(4 \, \pi \, a^2)$, 
the mean temperature and habitability tend to rise with 
increasing pressure (Fig. \ref{verticalCrossCut}).

\item
Remarkable differences in  surface temperature and habitability
exist between the low and high pressure regimes,
mainly because the range bracketed by extreme
surface temperatures decreases with increasing pressure (Fig. \ref{verticalCrossCut}).

\item
At low pressures ($p \lsim 0.3$\,bar) 
the  habitability is generally low and varies with $a$.
At high pressure ($p \gsim 1$\,bar), the  habitability is high and 
relatively constant inside the HZ. 


\item
In the temperature range suitable for
terrestrial mesophilic organisms ($0 \celsius \lsim T_m$ $\lsim 60 \celsius$),
the HZ moves away from the star as pressure increases,
rather than becoming broader.

\end{itemize}

The  characteristics of the  pressure-dependent  HZ 
result from the complex interaction of  physical effects
that   become stronger as the  surface pressure increases. 
The main ones are the intensity of the greenhouse effect
(recalling that we keep the atmospheric composition fixed),
the efficiency of latitudinal heat transport, 
and the broadening of the temperature range of liquid water. 
The increase of the greenhouse effect 
bends the contours of equal planet temperature towards higher distances from the star
as the pressure increases. 
The broadening of the liquid water range bends the inner edge of the
HZ closer to the star as pressure increases. 
The rise of the latitudinal heat transport  tends to yield   uniform
planet temperatures at high pressure. 

The comparison of our boundaries of habitability 
with the limits of the classic HZ  around a Sun-like star %
yields the following conclusions.  
Our inner edge, 
 calculated for Earth-like pressure, cloudiness and humidity,
is located at
$a=0.82$\,AU, in the range of the
inner limits 
predicted by Selsis et al. (2007). 
Our outer edge,   
 calculated for Earth-like atmospheric composition,
lies at $a=1.08$\,AU, 
much closer to the star than the quoted outer limits of the classic HZ,
 which are estimated for CO$_2$-rich planetary atmospheres
(Kasting et al. 1993, Selsis et al. 2007).

Thanks to the EBM capability of exploring
the latitudinal and seasonal variations of the surface temperature
we have found the following results:
\begin{itemize}
\item
The mean global planet temperature, $T_m$,
is not a good indicator of habitability at low pressure.
As an example, we find planets that are habitable
even if $T_m$ is well below the water freezing point (left panel of Fig. \ref{specificExamples}).
This result highlights the need of
solving the latitude profile to properly characterize
 planetary habitability. 
\item
Habitable planets
with  zonal temperatures above the water boiling point 
may in principle exist
(right panel of Fig. \ref{specificExamples}).
%
These cases are new candidate ``Class II'' habitats (Lammer et al. 2009),
i.e. bodies that become unsuitable to host life after an initial period of habitability.
\end{itemize}

Our results,
calculated for circular orbits, can   be  extended 
to planets in Keplerian orbits with moderate eccentricity ($e<0.5$), provided
one uses the effective circular semi-major axis
$a_\text{\em eff} = a (1-e^2)^{1/4}$ 
as a proxy of $a$. 
At higher eccentricities, however, the extent of the HZ shrinks
considerably (Fig. \ref{aeff_hab}). 
In principle, our results 
can  also be applied to planets 
orbiting stars different from the Sun, provided one uses the insolation
$q=L_\star/(4 \pi a^2)$ rather than the semi-major axis. 
In practice, there are several reasons, discussed in Section \ref{sectNonSolarStars},
why this type of generalization should be done with  caution.

The potential effects on climate  
of physical quantities unconstrained by observations
is a reason of concern in the study of exoplanets. 
Thanks to the flexibility of our EBM simulations 
we have explored how the habitability, and its dependence on pressure,
can be influenced by changes of rotation period, axis obliquity, planet geography, and surface albedo.
We find the following results:
\begin{itemize}
\item
The shapes of the curves of habitability $h$ versus $a$ 
are sensitive to variations of planetary parameters
at low pressure, but not at high pressure. 
\item
The general location and radial extension of the pressure-dependent habitable
zone are modestly influenced by planetary parameters.
\end{itemize}

In conclusion, climate EBMs  
offer a powerful tool to determine 
the range of stellar, orbital and planetary parameters 
that satisfy the liquid water criterion in exoplanets. 
The possibility of investigating  seasonal and latitudinal variations of the surface
temperature offers a significant insight on the habitability
and its long term evolution,
as well as
on the possible type of life that might exist
on the planetary surface. 
%
These type of studies will help to optimize the selection of targets
in future  searches for 
biomarkers in extrasolar planets. 
%

\acknowledgments

We acknowledge helpful conversations with
Marco Fulle, Michele Maris, Pierluigi Monaco, and Salvatore Scarpato. 
Thanks to Jost von Hardenberg for help in recovering and analysing Earth climate data.
We also thank Rodrigo Caballero and Raymond Pierrehumbert for
help on the use of their climate utilities.
The comments and suggestions
received from an anonymous referee have improved the presentation of this work.
JV thanks Kerrie for her enduring support and help in the English revision of the manuscript.


\appendix

\section{Model prescriptions \label{modelPrescriptions}}

%
We present  
the formalism adopted to model the physical quantities $C$, $D$, $I$, $A$, and $S$
that appear  in the diffusion equation (\ref{diffusionEq}).
General introductory remarks on these quantities
are given in  Section \ref{sectModel}. 
Further details can be found in the literature cited in the text.  
%

\subsection{Effective thermal capacity, $C$ \label{sectThermalCapacity}} 

The term $C$ 
represents the thermal inertia of the
atmospheric and surface layers that contribute to the surface energy budget. 
For a layer  with density $\rho$,  specific heat capacity $c_p$ and depth $\Delta \ell$, 
the effective thermal capacity is $C=\rho \, c_p \, \Delta \ell$. 
In a planet with oceans, the surface oceanic layers mixed by the winds provide
the main contribution to the thermal inertia. 
A wind-mixed ocean layer with $\Delta \ell= 50$\,m
has  $C_\text{ml50} = 210 \times 10^6$ J m$^{-2}$ K$^{-1}$ (WK97).
Also the contribution of the atmosphere can be expressed in terms of an equivalent mixed ocean layer.
For the Earth atmosphere, the equivalent ocean depth is of 2.4\,m (Pierrehumbert 2010),
corresponding to  $C_{\text{atm},\circ} = 10.1 \times 10^6$ J m$^{-2}$ K$^{-1}$. 
For other planetary atmospheres 
the thermal inertia will scale as
\begin{equation}
\left( { C_\text{atm} \over C_{\text{atm},\circ} } \right)
= \left( { c_p \over c_{p,\circ} } \right) \,
 \left( { p \over p_\circ} \right) ~~~,
 \label{Catm}
\end{equation}
where $c_p$ and $p$ are the specific heat capacity and total pressure of the atmosphere,
respectively (Pierrehumbert 2010).
The effective heat capacity of the solid surface is generally negligible.
As a representative value,  a layer of rock or ice with 
$\Delta \ell= 0.5$\,m yields
a contribution $C_\text{solid} \simeq 1 \times 10^6$ J m$^{-2}$ K$^{-1}$. 
This term, even if small, is not negligible
in planets without oceans and with very thin atmospheres. 
On the basis of the above considerations, we adopt 
\begin{eqnarray}
C_o= C_\text{ml50} + C_\text{atm}  \nonumber \\
C_l= C_\text{solid} + C_\text{atm} 
\label{CoceanCland}
\end{eqnarray}
for the thermal inertia of the atmosphere over oceans
and over lands, respectively. 
For ices not undergoing liquid-solid transition, we adopt $C_i=C_l$.
In the temperature range $263 < T < 273$,
we add an extra contribution to the thermal inertia of ice
to take into account the latent heat
of phase transition. In practice, based on similar recipes adopted by North et al. (1983)
and WK97, we adopt $C_i=C_l+C_\text{ml50}/5$. 
%

%

The mean effective thermal capacity  of each latitude zone  
is calculated using as weighting factors the coverage of oceans,
$f_o$, and continents, $f_l=1-f_o$. 
The expression that we use
\begin{equation}
C = f_l [(1-f_{i}) \, C_l+f_{i} \, C_{i}]    
+ f_o [ (1-f_{i}) C_o + f_{i} \, C_{i}] 
\label{ThermalCapacity}
\end{equation}
takes into account the ice
coverage,   $f_i$, estimated as explained in Section \ref{SectionCoverage}.

All together, our values of thermal capacity are
very similar to those adopted by WK97 and SMS08, with the exception
of $C_{\text{atm},\circ}$, which is about twice in our case.
Attempts to use higher values of the ocean capacity previously proposed by North et al. (1983)
and representative of a 75-m deep wind-mixed ocean layer,
yield a worse match between the model predictions and the Earth experimental data.


\subsection{Diffusion coefficient, $D$ \label{sect_D} \label{sectDiffCoefficient}}


%
The horizontal  transport of heat on the planet surface  
is mainly due to the general circulations and the related instabilities of the ocean and the atmosphere, which are governed by many factors. 
The most relevant are:
the physical and chemical properties of the atmosphere and oceans,  
the presence of Coriolis force 
induced by the planet rotation, and  the topography of the planet surface. 
Since EBMs are  zonally averaged, longitudinal variations of the heat transport are not considered. 
Even so, it is extremely difficult to model all the factors that govern the latitudinal transport.
To keep low the computational cost, EBMs incorporate 
the efficiency of the transport into a single quantity, namely
the diffusion coefficient $D$, 
even though representing latitudinal heat transport by a diffusive term is, in itself, a gross simplification.
In previous EBM work, this term 
has been parametrized using a scaling law that takes into account
the dependence of the diffusion coefficient 
on the main physical quantities involved in the latitudinal transport.
The relation proposed by WK97, 
\begin{equation}
\left( {D~ \over D_\circ} \right)  =  
\left( { p~ \over p_\circ} \right) \, \left( { c_\mathrm{p}~ \over c_{p,\circ}} \right) \,
\left( { m~ \over m_\circ} \right)^{\!\! -2} \, 
\left( { \Omega~ \over \Omega_\circ} \right)^{\!\! -2} ~~~,
\label{diffusionWK97}
\end{equation}
 gives $D$ as a function of
atmospheric pressure, $p$, specific heat capacity, $c_\mathrm{p}$, 
mean molecular weight, $m$, and angular  velocity of planet rotation, $\Omega$.
All these quantities are scaled with respect to the  reference Earth values. 
A discussion of the physics underlying this relation can be found in WK97.
A number of recent applications of EBM to exoplanet climates have adopted the same relation  
(SMS08, Dressing et al. 2010). 
We refer to  these papers 
for cautionary remarks regarding the limits of validity of the scaling relation $D \propto \Omega^{-2}$.

In our work, we address an intrinsic limitation of the above formulation, 
namely the fact that, once the planet parameters are specified, 
$D$ is constant in latitude and time.
This approximation is too crude because the Earth's latitudinal
transport is driven by atmospheric
convective cells characterized by a latitudinal pattern with seasonal variations. 
As a consequence, in real planets the efficiency of the diffusion will  vary in latitude and  time.  
%
Here we propose a formalism that introduces latitudinal and seasonal variations of $D$
related to the stellar zenith distance. This formalism can be applied to a generic planet. 

%
%
%

Most of the latitudinal heat transport is carried out by the  
circulation that takes place in the atmospheric convective cells. 
On the Earth, the most notable of these features are the two Hadley cells
that depart from the intertropical convergence zone (ITCZ) and produce
the net effect of a polar-ward transport. 
The position of the Hadley  cells shifts during the year,
influenced by the solar zenith distance.
This is demonstrated by the seasonal shift of the ITCZ, 
which moves to higher latitudes in the summer hemisphere.
Since the Hadley cells yield the largest contribution to the latitudinal heat transport,
we  conclude that 
the maximum efficiency of this type of transport 
is  larger when the zenith distance $Z_\star$ is smaller.
%
Inspired by this behavior,
and without making specific assumptions on the systems of atmospheric cells
that may be present
in a generic planet, we have introduced a dependence of 
the diffusion coefficient on $\mu=\cos Z_\star$.
More specifically, we assume that $D$ scales with
the mean diurnal value $\overline{\mu}$, derived from Eq. (\ref{muOverline}).
From this equation and Eq. (\ref{starDeclination}) one can see that
$\overline{\mu}$ depends on the latitude, $\varphi$, 
and the obliquity of the  axis of rotation, $\varepsilon$;
it also depends on the time of the year, $t$, which determines the seasonal value of $\delta_\star$ 
and $\lambda_\star$. 
Therefore, by  assuming a dependence $D=D(\overline{\mu})$ we 
introduce, in practice, a dependence on  $\varphi$, $t$ and $\varepsilon$.
To incorporate this effect in the EBM prescriptions,
we adopt the following expression 
%
%
%
\begin{equation}
\left( {D~ \, \over D_\circ} \right)  = \zeta_\varepsilon(\varphi,t) \, 
\left( { p~ \, \over p_\circ} \right) \, \left( { c_\mathrm{p}~ \over c_{p,\circ}} \right) \,
\left( { m~ \, \over m_\circ} \right)^{\!\! -2} \, \left( { \Omega~ \, \over \Omega_\circ} \right)^{\!\! -2} ~~~,
\label{ModulatedDiffusion1}
\end{equation}
where $\zeta_\varepsilon(\varphi,t)$ is a modulating factor
that scales linearly with $\overline{\mu}$. This factor is
normalized in such a way that the mean global annual value of $D$ equals that given
by the scaling law (\ref{diffusionWK97}). 
Thanks to this fact, we can make use of the previous studies of  the diffusion coefficient,
as far as the dependence on $p$, $c_\mathrm{p}$, $m$, $\Omega$ is concerned,
and of our own formalism, as far as the dependence on $\varphi$, $t$ and $\varepsilon$
is concerned. 
We now show how
the modulating factor $\zeta_\varepsilon(\varphi,t)$
can be expressed as a function of a single parameter, $\mathcal{R}$,
that represents the ratio between the maximum and minimum value of the diffusion
coefficient in the planet. 
%

\subsubsection{The modulating factor $\zeta$}

To obtain an expression for the modulating factor, $\zeta$, 
we use the following defining conditions:
(i) $\zeta$ scales linearly with $\overline{\mu}$;
(ii)
$\zeta$ is normalized in such a way that its mean global annual value  is
$\leftmean \zeta_\varepsilon(\varphi,t) \rightmean = 1$.
%
The first  condition translates into
\begin{equation}
\zeta_\varepsilon(\varphi,t) = c_0 + c_1 \, \overline{\mu}_\varepsilon(\varphi,t)   ~~~,
\label{ModulationFactor1}
\end{equation}
where $c_0$ e $c_1$ are two constants.
The second condition becomes
\begin{equation}
\leftmean \zeta_\varepsilon(\varphi,t) \rightmean =
c_0 + c_1 \, \leftmean \overline{\mu}_\varepsilon(\varphi,t) \rightmean
= 1 ~~~,
\label{condition2}
\end{equation}  
where
\begin{equation}
\leftmean \overline{\mu}_\varepsilon(\varphi,t) \rightmean =
{
\int_0^P dt \int_{-{\pi \over 2}}^{+{\pi \over 2}} d\varphi ~ \cos \varphi ~\overline{\mu}_\varepsilon(\varphi,t)  
\over
\int_0^P dt \int_{-{\pi \over 2}}^{+{\pi \over 2}} d\varphi ~ \cos \varphi 
}   ~~~.
\label{doubleintegral}
\end{equation}

To determine $c_0$ and $c_1$, we need an additional relation between 
these two constants.
For this purpose, we follow an approach similar to North et al. (1983), who adjusted the diffusion coefficient
to be 3 times as large at the equator as it is at the poles.
In the framework of our formulation, this condition would be expressed in the form 
$\mathcal{R}=3$, where
\begin{equation}
\mathcal{R}={ [\zeta_\varepsilon(\varphi,t)]_{\max} \over [\zeta_\varepsilon(\varphi,t)]_{\min} }
\label{diffusionRatio}
\end{equation}
is the ratio between the maximum and minimum value of the modulation factor
in any latitude zone and at any orbital phase.  
The above condition is equivalent to
\begin{equation}
\mathcal{R}={ c_0 + c_1 \, [\overline{\mu}_\varepsilon(\varphi,t)]_{\max} 
\over c_0 + c_1 \, [\overline{\mu}_\varepsilon(\varphi,t)]_{\min} }
=
{ c_0 + c_1 \, [\overline{\mu}_\varepsilon(\varphi,t)]_{\max} 
\over c_0  }  ~~~,
\label{conditionR}
\end{equation}
where we have used the fact that $[\overline{\mu}_\varepsilon(\varphi,t)]_{\min}=0$
since $\overline{\mu}=0$ in the zones where the star lies below the horizon during a period of rotation. 
We treat $\mathcal{R}$ as an input parameter of the model used
to estimate $c_0$ and $c_1$.
By combining Eqs. (\ref{condition2}) and (\ref{conditionR}) it is easy to show that
\begin{equation}
c_1 = 
\left( 
{ [\overline{\mu}_\varepsilon(\varphi,t)]_{\max}  \over \mathcal{R}-1 } + \leftmean \overline{\mu}_\varepsilon(\varphi,t)\rightmean 
\right)^{-1}
\end{equation}
and
\begin{equation}
c_0 = c_1 \times
{ [\overline{\mu}_\varepsilon(\varphi,t)]_{\max}  \over \mathcal{R}-1 }   ~~~.
\end{equation}

The quantities $[\overline{\mu}_\varepsilon(\varphi,t)]_{\max}$
and $\leftmean \overline{\mu}_\varepsilon(\varphi,t)\rightmean$ 
are calculated  given the obliquity $\varepsilon$.
To compute the double integral in Eq. (\ref{doubleintegral})
we use the expression
of $\overline{\mu}$ given in Eq. (\ref{muOverline}),
with the  condition $\overline{\mu}=0$ when $H \leq 0$, i.e. when the Sun is below the local horizon
for a complete rotation of the planet. 
%

\subsubsection{Calibration of $\mathcal{R}$}

The parameter $\mathcal{R}$ was tuned to improve
the match between the observed and predicted
temperature-latitude profile of the Earth. 
The observed profile shows two features which are quite difficult to 
reproduce with a simple EBM.
One is the almost flat temperature profile within the tropical belt, 
the other is the sharp temperature drop at latitude $\varphi \simeq -60^\circ$
(Figs. \ref{earthLatTemp} and \ref{earthSeasons}).
By increasing $\mathcal{R}$ we are able to improve
the match to both features. 
We obtain the best results by adopting $\mathcal{R}=6$.
%
The temperature profile that we obtain is flatter at the tropics
than in previous work (Fig. \ref{earthLatTemp});
the temperature drop at $\varphi \simeq -60^\circ$ is reproduced
particularly well for the months of June, July and August
(see the case of July in Fig. \ref{earthSeasons}). 
The disagreement between the observed and modeled  profiles
at $\varphi < -60^\circ$ is probably due to the particular characteristics
(e.g. the high surface elevation) of Antarctica.

%
%
%

\subsection{The outgoing longwavelength radiation, $I$ \label{sectOLRcal}}

In climate EBMs,  the OLR is usually 
expressed as a function of the temperature.
In our problem, we are interested in probing variations of total
pressure and therefore we need an expression $I = I(T,p)$.
The pressure dependence of the OLR
enters through the infrared optical depth of the atmosphere,
$\tau_\text{IR}$,
which governs the intensity of the greenhouse effect.   
In principle,
one may expect a strong dependence of $\tau_\text{IR}$ on $p$  
as a result of two effects.
First, in a planetary atmosphere  
$\tau_\mathrm{IR} = \kappa_\mathrm{IR} \, p/g$,
where $\kappa_\mathrm{IR}$ is the  total absorption coefficient
and $g$ the gravitational acceleration. 
%
Second, in the range of planetary surface pressures, $\kappa_\mathrm{IR}$ 
is dominated by collisional broadening, which introduces a linear dependence
on $p$ of the widths of the absorption lines 
 (Pierrehumbert 2010; Salby 2012).
In absence of line saturation, 
$\kappa_\mathrm{IR}$ should therefore increase linearly with $p$. 
%
All together, one might expect 
$\tau_\mathrm{IR} = \kappa_\mathrm{IR} \, p/g \propto p^2$.
In practice, however, the rise of the optical depth is milder
because (1) the absorptions of different species may overlap in wavelength,
(2) lines are partly saturated, and 
(3) the amount of water vapor, which is an important contributor to $\kappa_\mathrm{IR}$,
is independent of $p$.
Given the complex interplay of these factors, it is not possible to derive
a simple analytical function $\tau_\mathrm{IR}=\tau_\mathrm{IR}(T,p)$.
%
Therefore, in order to obtain $I=I(T,p)$,
we performed a series of radiative calculations and tabulated the results as
a function  of $T$ and $p$.

%
%
We used  standard radiation 
models developed at the National Center for Atmospheric Research
(NCAR), as part of the Community Climate Model (CCM) project. 
The calculations were performed in a column radiation model scheme
for a cloud-free atmosphere (Kiehl \& Briegleb, 1992).
We used the set of routines CliMT (Pierrehumbert 2010, Caballero 2012),
adopting the Earth's value of surface gravitational acceleration. 
We varied $p$ while keeping 
the mixing ratios of  non-condensable greenhouses gases (CO$_2$ and CH$_4$)
equal to the Earth's values.
The reference values of total pressure and partial pressure
of greenhouse gases are shown in Tables \ref{tabEarthData} and \ref{tabCalibratedPar}. 
The contribution of water vapor was parametrized through the relative humidity, $\rh$,
as we explain below. 
In the regime of high temperatures, 
we increased the resolution of the vertical strata (up to 10000),
in order to better track the water vapor in the higher atmospheric levels.
To test the radiative calculations in the 
regime of very low pressure (i.e. negligible atmospheric greenhouse)
we compared the predicted OLR with the black body radiation
calculated at the planet surface. 
As expected, the results are identical as long as water vapor is negligible.

\begin{figure}
\begin{center}
\includegraphics[width=8cm]{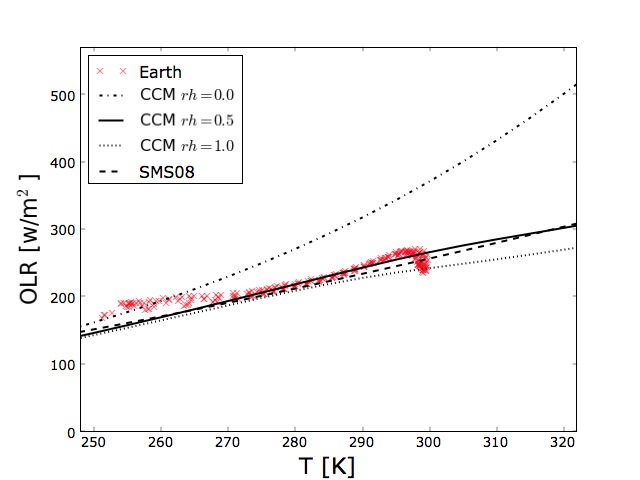}
\caption{
Comparison of radiative calculations and 
experimental data of the outgoing long-wavelength radiation (OLR)
of the Earth.
Crosses: mean annual data obtained from the ERBE satellite for
the years 1985-1988.
CCM radiative calculations for an Earth-type atmosphere
with relative humidity $\rh$=0, 0.5, and 1.0
are shown as dot-dashed, solid and dotted curves, respectively,
after subtraction of the mean global long-wavelength cloud forcing
(see Section \ref{sectOLRcal}).
Dashed curve: OLR model adopted by SMS08.
\label{figOLR}}
 \end{center}
\end{figure}

To calibrate the OLR for the case of the Earth,
we used as a reference the mean annual  Earth's OLR 
measured with the ERBE satellite
(online material published by Pierrehumbert 2010). 
In practice, 
we combined two ERBE data sets --- i.e.,
the OLR versus latitude and the surface temperature versus latitude ---
to build a set of  OLR data  versus temperature.
These experimental data are shown
with crosses in  Fig.\,\ref{figOLR}. 
To test different types of  models, we then used the CCM 
to calculate the clear-sky OLR as a function of temperature 
for different values of
relative humidity ($rh=0.1$, 0.5 and 1.0).
%
We then subtracted 28 W m$^{-2}$ to the clear-sky results to take into account 
the 
mean global  long-wavelength forcing of the clouds on the Earth  (Pierrehumbert 2010).
The  three CCM curves  
resulting for $rh=0.1$, 0.5 and 1.0
are shown in Fig. \ref{figOLR}.
The curve obtained for $rh=0.5$ (solid line) gives the best match to the experimental data 
and is very similar to the OLR adopted by SMS08 and SMS09 (dashed line).
Based on these results, we adopted $rh=0.5$ in our model.  
For applications to planets with cloud coverage $\leftmean f_c \rightmean$
different from that of the Earth, $\leftmean f_{c,\circ} \rightmean$,
we subtract $28 \, (\leftmean f_c \rightmean/\leftmean f_{c,\circ} \rightmean)$ W\,m$^{-2}$
to the clear-sky OLR obtained from the radiative calculations.

\subsection{Albedo, $A$ \label{sectionAlbedo}}

In the formulation of SMS08, the albedo is an analytical function of the temperature
which, in practice, considers two types of surfaces: with ice and without ice.
%
In our work, we employ a formalism  that 
can be applied to planets with any type of surface characteristics. 
%
%
For each latitude zone 
we calculate a mean albedo 
%
  by weighting
the  contributions  of continents, $a_l$,
oceans, $a_o$,   clouds, $a_c$, ice on continents, $a_{il}$, and ice on oceans, $a_{io}$.
The weighting factors are the zonal coverage of oceans,
$f_o$, and continents, $f_l$;
within oceans and continents we  separate the contribution of ice, $f_i$;
the cloud coverage on  water, land and ice is specified by the parameters
$f_{cw}$, $f_{cl}$ and $f_{ci}$, respectively. 
In this way we obtain        
\begin{eqnarray}
\lefteqn{
a_s=  f_o \Big\{ (1-f_{i}) \Big[ a_{o} (1-f_{cw}) + a_{c}f_{cw} \Big] +  {} }  \nonumber\\
& &  {}   + f_{i}   \Big[ a_{io} (1-f_{ci}) + a_{c} f_{ci} \Big] \Big\}+  \nonumber\\
& &  {}   +     f_l \Big\{ (1-f_{i})  \Big[ a_{l} (1-f_{cl}) + a_{c} f_{cl} \Big] + \nonumber\\
& &  {}   + f_{i} \Big[ a_{il} (1-f_{ci}) + a_{c} f_{ci} \Big] \Big\} ~.
\label{surfaceAlbedo}
\end{eqnarray}
In this equation we have omitted for simplicity the latitude dependence
of the zonal coverage of oceans and lands. 
This formulation of the surface albedo is similar to that adopted by WK97.
At variance with that work,  we consider the ice on continents (not only on oceans)
and variable cloud cover.
Only a few parameters in the above expression are free. 
Most of the albedo parameters 
are estimated with specific prescriptions that we describe below.
The zonal coverages of oceans, lands, ice and clouds are described in Section \ref{SectionCoverage}. 
In practice,  
only the zonal fraction of oceans, $f_o$, and the continental albedo, $a_l$,
are free parameters. 


%
%

 %

\subsubsection{Albedo parameters}

For the albedo of the continents, we adopt
$a_l=0.2$, the same value used by WK97, as a representative
value of the Earth continents. For other planets we treat $a_l$ as a free
parameter that can typically vary between 0.1 and 0.35  
(see Section \ref{sectAlbedoLands}).

The albedo of the oceans plays a crucial role in the estimate of $a_s$.
The ocean reflection depends on the zenith distance of the star, $Z_\star$, that
the EBM provides  self-consistently as a function of time and latitude. 
WK97  adopted the Fresnel formula (Kondratyev 1969)
to model the ocean reflectivity. However, the Fresnel formula is an ideal approximation
valid for smooth surfaces while the actual sea level is never ideally smooth.
For this reason we opted for a  function calibrated with experimental data
(Briegleb et al. 1986; Enomoto 2007):
\begin{eqnarray}
\lefteqn{
a_o = { 0.026 \over (1.1 \, \mu^{1.7} + 0.065)} + {} }  \nonumber\\
& &  {}  + 0.15 (\mu-0.1) \, (\mu-0.5) \, (\mu-1.0)   ~,
\end{eqnarray}
where $\mu = \cos Z_\star$ is calculated with Eq. (\ref{muDefinition}). 
Compared to the Fresnel formula, this expression gives very similar results
at $Z_\star \lsim 40^\circ$, but a less steep, more realistic, rise at high values of zenith distance 
(Enomoto 2007).

%
For the albedo of ice over lands and over ocean we adopted
$a_{il}=0.85$ and $a_{io}=0.62$, respectively. 
The existence of a difference between these two types of albedo 
has been documented in previous work (Kondratyev 1969, Pierrehumbert 2010).
By adopting  a high albedo for ice over continents we improve
the modelization of the Earth climate on the Antarctica,
which is a critical point of EBMs (see Fig. \ref{earthLatTemp} and 
Section \ref{modelCalibration}).

The cloud albedo, $a_c$,  
depends on  the microphysical properties and geometrical dimensions of the clouds,
which determine their optical properties.
%
For a cloud of given optical thickness, the albedo increases
with increasing stellar zenith angle, $Z_\star$, which elongates the slant optical depth (Salby 2012).
%
%
A linear dependence of $a_c$ on $Z_\star$ was reported by Cess (1976)
for a  set of  data 
representative of the Earth cloud albedos.
%
On the basis of that result, WK97 adopted 
$a_c =  \alpha + \beta Z_\star$ and 
tuned the parameters $\alpha$ and $\beta$ in such a way to match the total Earth albedo 
with their model. 
Since this tuning yields $\alpha<0$, a problem with this formalism is that 
the cloud albedo vanishes at low zenith distances and
becomes negative at $Z_\star=0$. 
We expect, instead, a minimum value of cloud albedo
to exist at $Z_\star=0$, corresponding to the minimum vertical thickness of the cloud.
For this reason, in our work we adopt the same parametrization, but with 
a minimum value of cloud albedo at low zenith distances, $a_{c0}$.
In practice, we use the expression 
\begin{equation}
a_c =  
\max \left\{  a_{c0} ,
\left[ \alpha +  \beta  Z_\star  \right] 
\right\} ~~~.
\label{cloudAlbedo}
\end{equation} 
%
We estimate $\alpha$ and $\beta$ 
by requiring the relation
$\alpha + \beta Z_\star$ to yield a good fit to Cess data. 
We  then tune  $a_{c0}$  in such a way to
match the total Earth albedo with our model.
In this way we obtained
$\alpha=-0.07$, $\beta=8 \times 10^{-3} (^\circ)^{-1}$,
and $a_{c0}=0.19$. 

\subsection{Incoming stellar radiation, $S$ \label{sectISR}}

The incoming stellar radiation is calculated as
the diurnal average 
\begin{equation}
S=  { \int_{0}^{2\pi} s(\hourangle) \, d\hourangle \over  \int_{0}^{2\pi} d\hourangle }
\label{meanDiurnalFluxDef}
\end{equation}
of the stellar flux $s(\hourangle)$ incident 
on the planet
at latitude $\varphi$, where 
$\hourangle$ is the instantaneous hour angle of the  star measured in angle units
from the local meridian.
To estimate $s(\hourangle)$ we proceed as follows. 
 %
%
At the time $t$ 
the planet is located at a distance $r=r(t)$ from its star. 
The stellar flux   
at such distance is 
\begin{equation} 
q(r) = { L_{\star} \over 4 \pi r^2 } ~,
\end{equation}
where $L_\star$ is the bolometric luminosity of the star.
If we call $Z_\star$ the stellar zenith angle,
the instantaneous flux  at the planet surface is
\begin{equation}
s(\hourangle) =  
\begin{cases}
\tau_a \,  q(r) \, \cos Z_\star  & \text{if $|\hourangle| < H$}  \\
0 & \text{if $|\hourangle| \geq H$} 
\label{instantFlux}
\end{cases}
\end{equation} 
where  
$\tau_a$ is the short wavelength
transmissivity of the atmosphere
and 
$H$ is the half-day length.
By definition, 
$-H \leq \hourangle \leq +H$ represents the portion of  rotation period
during which the star stays above the horizon.
The half-day length
in radians can be estimated from the expression (WK97)
\begin{equation}
\cos H = -\tan \varphi \tan \delta ~~~~~(0 < H < \pi)~.
\label{halfDay}
\end{equation}
The flux $s(\hourangle)$ is  null when $|\hourangle| \geq H$
because in this case the star is below the horizon ($Z_\star \geq \pi/2$).  
%
%
The zenith distance, $Z_\star$, is related to the latitude, $\varphi$,  the stellar declination, $\delta_\star$,
and the hour angle, $\hourangle$, by means of the equation 
\begin{equation}
\cos Z_\star = \sin \varphi \, \sin \delta_\star + \cos \varphi \, \cos \delta_\star \, \cos \hourangle ~.
\label{muDefinition}
\end{equation}

With the above relations, we can calculate the average (\ref{meanDiurnalFluxDef})
along circles of constant latitude. In these circles, the terms
$\sin \varphi$ and $\cos \varphi$ are constant in the integration. 
Also the  declination $\delta_\star=\delta_\star(t)$ and the 
flux  $q(r)=q(r[t])$ can be treated as constants
if the rotation period is much smaller than the orbital period, i.e. if
\begin{equation}
P_\mathrm{rot} \ll P_\mathrm{orb} ~.
\end{equation}
With these assumptions, if the  atmosphere
is transparent in the short wavelength  range ($\tau_a=1$),
it is easy to show that
\begin{equation}
S= { q(r) \over \pi} 
\left( H\sin \varphi \, \sin \delta_\star + \cos \varphi \, \cos \delta_\star \, \sin H \right) ~.
\label{meanDiurnalFlux}
\end{equation}

In a similar fashion, by defining $\mu = \cos Z_\star$, it is possible to calculate the mean diurnal value of $\mu$
when the star is above the horizon, 
$\overline{\mu}=\int_{-H}^{H} \mu \, d\hourangle/\int_{-H}^{H} d\hourangle$,
and obtain the relation
\begin{equation}
\overline{\mu} = \sin \varphi \, \sin \delta_\star + \cos \varphi \, \cos \delta_\star \,  { \sin H \over H } ~
\label{muOverline}
\end{equation}
that we use in our formulation of the diffusion coefficient. 
 
Once we have obtained the diurnal average $S$, we calculate its temporal variation
in the course of the orbital period, $S=S(t)$.
The time $t$ specifies the planet position along the orbit and
$S=S(t)$ represents the seasonal evolution of the flux at a given latitude.
The only quantities that depend on $t$ in Eq. (\ref{meanDiurnalFlux}) are
$q(r)$ and $\delta_\star$. 
To calculate  the seasonal evolution of $q(r)$ and $\delta_\star$ we proceed as follows. 

\begin{figure}[]   
\includegraphics[width=8.cm]{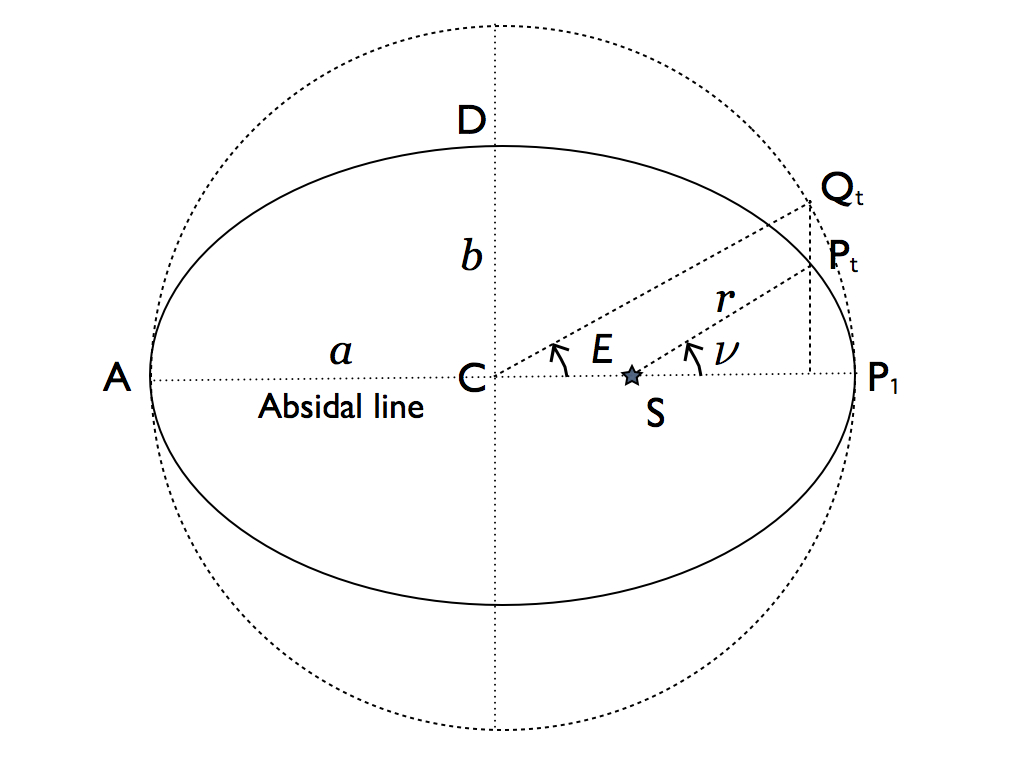} 
\includegraphics[width=8.cm]{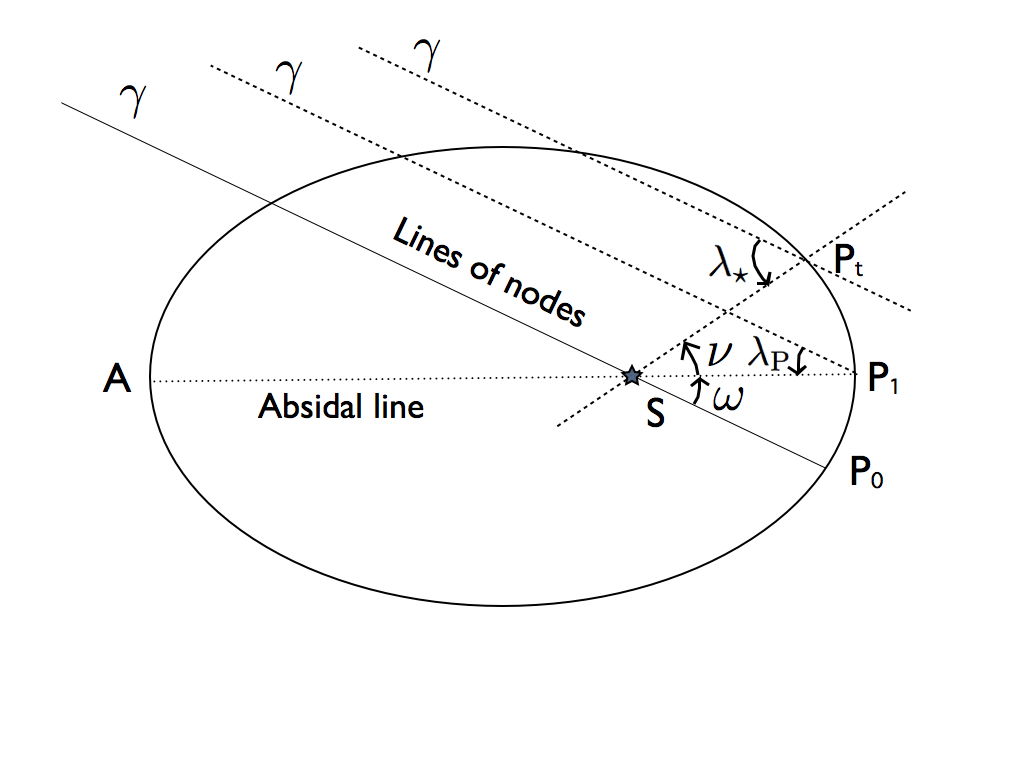} 
\caption{ 
Orbital elements used in the derivation
of the incoming stellar radiation (Section \ref{sectISR}). 
The ellipsis represents the planet orbit;
S is the star, located at one focus of the ellipsis;
P$_t$  the position of the planet at  time $t$;
P$_1$  the planet at the pericenter;
the absidal line  connects the pericenter to the apocenter, A.
The semi-major axis is $a=$AC; the semi-minor axis $b=$CD$=a \sqrt{(1-e^2)}$.
Left panel:   elements used to demonstrate relations (\ref{r_a}) and (\ref{nu_keplerE}); 
$E$ is the {\em eccentric anomaly};
Q$_t$ the projection of P$_t$ on the circle that circumscribes the orbit. 
Right panel: demonstration of the relation (\ref{Ls_nu})
between the planetocentric orbital longitude of the star, $\lambda_\star$,
the true anomaly, $\nu$, and the argument of the pericenter, $\omega$. 
The line of nodes is the intersection between the  orbital plane
and the equatorial plane of the planet; P$_0$ is the planet 
at the ascending node. 
For each position of the planet we show the instant direction of the lines of nodes, $\gamma$.
  }
 \label{orbitFig}%
\end{figure}

We specify the position of the planet on its orbit  
using a system of polar coordinates centered on the star
and origin of the angles in the direction to the pericenter, as shown in Fig. \ref{orbitFig}.
The  position of the planet at time $t$, P$_t$,
is specified  by $r=r(t)$ and the {\em true anomaly}, $\nu=\nu(t)$.
The instant stellar flux is given by 
$q (r) = { q_0 \over (r/a)^2 }$ where
 $q_0 = L_{\star} / (4 \pi a^2)$. 
The instantaneous value of $r/a$ can be calculated 
by introducing an additional angular variable $E$ (the {\em eccentric anomaly})
such that  
\begin{equation}
r/a = 1-e \cos E ~~~,
\label{r_a}
\end{equation}
where $e$ is the eccentricity and
$E$ is interpreted geometrically in the left panel of  Fig. \ref{orbitFig}.
We show below how to compute $E$. 

To calculate the stellar declination $\delta_\star$
we consider the {\em planetocentric orbital longitude of the star}, 
$\lambda_\star$, shown in the right panel of  Fig. \ref{orbitFig}.
The origin of this angle is the line of nodes,
defined by the intersection of the orbital plane with the equatorial plane.  
Without loss of generality, we set $t=0$ the instant in which the planet crosses the
ascending node\footnote
{
The intersection of the orbital plane with a reference plane 
used to measure the orbit inclination
is called the ``lines of nodes''. 
Here the adopted reference plane is the equatorial plane 
and the inclination coincides with the axis obliquity. 
The planet orbit intersects the lines of nodes
in two points, called the ascending and descending node.
The ascending node is the one in which $d (\delta_\star)/dt > 0$. 
}.
With this choice, the instant value of the stellar declination  is (Allison \& McEwen 2000)
\begin{equation}
\label{starDeclination}
\delta_\star = \arcsin \left( \sin \varepsilon \, \sin \lambda_\star  \right)  ~~~,
\end{equation}
where $\varepsilon$ is the obliquity, i.e. the inclination of the planet's orbit to its equator. 
%
To calculate $\lambda_\star(t) $ we consider its relation with the true anomaly, $\nu$.
From the top panel of  Fig. \ref{orbitFig} it is easy to see that  
\begin{equation}
\lambda_\star(t) = \nu(t)+\lambda_\mathrm{P} ~~~,
\label{Ls_nu}
\end{equation}
where $\lambda_\mathrm{P}$ is the planetocentric  longitude of the star
at the moment in which the planet is at the pericenter (P$_1$ in the figure);
in terms of orbital parameters, 
$\lambda_\mathrm{P}=\omega$, where $\omega$ is the {\em argument of the pericenter}.
%
To solve the  equation  (\ref{Ls_nu}), we use 
the expression  [BF90 (Eq. 10.30, p. 211)]
\begin{equation}
\tan (\nu/2) = \left( { 1+e \over 1-e} \right)^{1 \over 2} \tan (E/2)
\label{nu_keplerE}
\end{equation}
that can be inferred from the top panel of Fig. \ref{orbitFig}.  

At this point we are left with the calculation of the eccentric anomaly $E$. 
To this end we use Kepler's equation 
\begin{equation}
E-e \sin E = M ~~~,
\label{KeplerEquation}
\end{equation}
where $M$ is the {\em mean anomaly}, defined to be a linear
function of time which increases by $2\pi$ per revolution according to  the expression
\begin{equation}
M= \overline{n} \, t + M_\circ ~~~,
\label{MeanAnomaly}
\end{equation}
where $\overline{n}=2\pi/P$.
The constant $M_0$ is determined by the choice of the initial conditions.  
%
%
%
%
To solve the transcendent Kepler's equation (\ref{KeplerEquation}) 
we use Newton's iteration method\footnote{
Newton's method is based on the iteration 
$x_{i+1} = x_i - f(x_i)/f'(x_i)$.
In practice, in our case  $f(x)=E-e\sin E-M$ and therefore
$$
E_{i+1}=E_i - { E_i - e \sin E_i - M \over 1 - e \cos E_i} ~~~.
$$
}.

%
%
%



\subsection{Zonal coverage of oceans, lands, ice and clouds \label{SectionCoverage}}


The zonal ocean fraction, $f_o$, is a free parameter that also determines
the fraction of continents $f_l=1-f_o$.
By assigning $f_o$ and $f_l$  to each latitude zone of the planet, the EBM takes
into account the planet geography, although in a schematic way.



The zonal coverage of ices, $f_i$, is calculated
as a function of the  mean diurnal zonal  temperature, $T$. 
This approach, adopted by WK97 and followed by SM08, allows to incorporate 
the climate feedback between temperature and ice albedo
into the EBM. 
In their original formulation,  WK97   
used data from 
Thompson and Barron (1981) 
to calculate the ocean fraction covered by sea ice as  
\begin{equation}
f_i (T)=  \max 
\left\{ 0, 
 \left[ 1 - e^{ (T-273\,\mathrm{K} )/ 10\,\mathrm{K} }
 \right]  \right\} ~~~.
\label{freezingCurve}
\end{equation} 
This expression provides a  ``freezing curve'' 
that ideally describes
the formation of ices as the temperature decreases. 
However, in this formulation  the ice melts completely and instantaneously
as soon as  $T > 273$\,K.
This approximation is too crude for zones that are frozen for most of the time:
in such regions  we expect a build-up of permanent ices if the time-scales
for ice formation and ice melting are comparable. 
To avoid this problem, we compare the time intervals
in which $T < 273$\,K and $T \geq 273$\,K in each zone. 
 If  $T < 273$\,K during more than 50\% of the orbital period, 
we adopt a constant ice coverage for the full orbit, $f_i=f_i (\overline{T})$,
where $\overline{T}$ is the mean {\em annual} zonal temperature.
In the other cases, we follow the original formulation 
(\ref{freezingCurve}). 
In this way we obtain the formation of permanent ices, with a coverage
that increases as the mean zonal annual temperature decreases.
%
Our treatment significantly improves the description 
of the  ice coverage on the Earth, avoiding a sudden appearance and
disappearance of the polar caps in the course of each orbit. 
%
%
%
%
To keep  the model  simple, we adopt for the continents the 
same ice coverage of the oceans. 
A more realistic treatment of the ice coverage would require a formulation of 
the  formation and melting of ices as a function of time. 
The dynamical treatment of the ice formation and melting 
is beyond the purpose of this paper
and will be the subject of subsequent work. 
e refer to Spiegel et al. (2010) for a detailed discussion of ice formation and melting
using an EBM.


As far as the cloud coverage is concerned,
 we adopt
different values for clouds  over oceans, continents and ices
in our formulation of the albedo [Eq. (\ref{surfaceAlbedo})].
The observational support for variations of the cloud coverage
on different types of underlying surfaces comes  from 
an analysis of  Earth   data
performed by Sanrom\'a \& Pall\'e (2012). 
These authors used  data 
collected by  the International Satellite Cloud Climatology Project
and presented the cloud coverage as a function of latitude for clouds on
water, ice, desert and vegetation.  We averaged these latitude profiles
to obtain a global representative cloud coverage for each type of surface.
%
%
The value for the water was slightly adjusted to improve the agreement of our model 
with the experimental albedo-latitude profiles of the Earth. 
The cloud coverage over continents was calculated
by weighting deserts and vegetation with  factors 2/3 and 1/3, respectively\footnote{
http://phl.upr.edu/library/notes/ vegetationiceanddesertsofthepaleo-earth.}.
As a result, we adopt 
$f_{cw}=0.67$, $f_{cl}=0.50$ and  $f_{ci}= 0.50$
for clouds over water, continents, and ices, respectively. 
With this choice of parameters we obtain a global cloud coverage 
$\leftmean f_c \rightmean=0.612$ in our best model for the Earth.
This value is in  agreement
with the experimental  value $\leftmean f_c \rightmean=0.603$ obtained from  
the  ERA-Interim reanalysis for the years 1979-2010 
(Dee et al. 2011). 
For comparison, WK97 adopted $\leftmean f_c \rightmean = 0.5$
in their formulation of the surface albedo of the Earth. 
An interesting feature of our formalism is that the cloud
coverage is automatically adjusted for planets with different 
fractions of continents, oceans and ices.

{}

\clearpage

%
%

\clearpage
 
\begin{deluxetable}{cll}
\tabletypesize{\scriptsize}
\tablecaption{Earth data}
\tablewidth{0pt}
\tablehead{\colhead{Parameter} & \colhead{Value} & \colhead{Comment} }
\startdata 
$q_0$ & 1361.6 W m$^{-2}$  & Solar constant\tablenotemark{a}    \\ 
$a$  & 1.000 AU & Semi-major axis \\
$e$  & 0.01671022   & Orbital eccentricity \\
$\epsilon$ & 23.43929 & Obliquity \\
$A_{m,\circ}$ & $0.328$  &  Mean annual global albedo\tablenotemark{b} \\
$T_{m,\circ}$ & $287.44$\,K &  Mean annual global surface temperature\tablenotemark{c}  \\
$p_{t,\circ}$ & $1.0132 \times 10^5$\,Pa  &  Total Earth surface pressure \\ 
CO$_2$ & 380 ppmV & Volumetric mixing ratio of CO$_2$\\
CH$_4$ & 1.7 ppmV & Volumetric mixing ratio  of CH$_4$\\
$c_{p,\circ}$ & $1.005 \times 10^3$ J Kg$^{-1}$ K$^{-1}$ & Specific heat capacity of the atmosphere \\
$m_\circ$ & 28.97 & Mean molecular weight of the atmosphere \\
\enddata
\tablenotetext{a}{From Kopp \& Lean (2010). The measurement given in that paper,
$1360.8 \pm 0.5$ W m$^{-2}$, was obtained during the 2008 solar minimum. 
The excursion between solar minimum and maximum
quoted in the same paper amounts to $1.6$ W m$^{-2}$.
We have added half this excursion to the  value measured at the minimum.}
\tablenotetext{b}{Area-weighted mean annual albedo of the Earth
measured from the average ERBE data for the period 1985-1989 (data taken from courseware of Pierrehumbert 2010).}
\tablenotetext{c}{Area-weighted mean annual surface temperature of the Earth
measured from ERA Interim data for the years 1979-2010 (see Dee et al. 2011) }
\label{tabEarthData}
\end{deluxetable}

\begin{deluxetable}{cllll}
\tabletypesize{\scriptsize}
\tablecaption{Fiducial model parameters}
\tablewidth{0pt}
\tablehead{\colhead{Parameter} & \colhead{Fiducial value} & \colhead{Comment} 
& \colhead{Eq.} & \colhead{Source} }
\startdata
$C_{\text{atm},\circ}$ & $10.1 \times 10^6$ J m$^{-2}$ K$^{-1}$ & Effective thermal capacity of the Earth atmosphere &
(\ref{Catm}) & Pierrehumbert (2010) \\
$C_\text{ml50}$ & $210 \times  10^6$ J m$^{-2}$ K$^{-1}$ & Effective thermal capacity of the oceans &
(\ref{CoceanCland}) & WK97 \\ 
$C_\text{solid}$ & $1 \times 10^6$ J m$^{-2}$ K$^{-1}$ & Effective thermal capacity of the solid surface &
 (\ref{CoceanCland}) & This work \\ 
$D_\circ$ & 0.600 W m$^{-2}$ K$^{-1}$ & Diffusion coefficient  &
(\ref{ModulatedDiffusion1}) & Pierrehumbert (2010) \\
$\mathcal{R}$ & 6 & Maximum excursion of diffusion efficiency &
(\ref{diffusionRatio})
& This work \\
$a_l$ & 0.20      & Surface albedo of lands &  (\ref{surfaceAlbedo}) & WK97 \\
$a_{il}$ & 0.85     & Surface albedo of ice on lands & (\ref{surfaceAlbedo})  & 
Pierrehumbert (2010)\\ 
$a_{io}$ & 0.62      & Surface albedo of ice on ocean & (\ref{surfaceAlbedo}) & 
This work \\ 
$f_{cw}$ & 0.67           &  Cloud coverage on water & (\ref{surfaceAlbedo}) &
This work \\ 
$f_{cl}$ & 0.50             & Cloud coverage on land & (\ref{surfaceAlbedo}) &
This work \\ 
$f_{ci}$ & 0.50             & Cloud coverage on ice & (\ref{surfaceAlbedo}) & 
This work \\ 
\enddata 
\label{tabCalibratedPar}
\end{deluxetable}

\begin{deluxetable}{cll}
\tabletypesize{\scriptsize}
\tablecaption{Model parameters fixed in the simulations} 
\tablewidth{0pt}
\tablehead{\colhead{Parameter} & \colhead{Adopted value} & \colhead{Comment}   }
\startdata
$N$  & 37 & Number of latitude zones \\
$L_\star$ & 
$L_{\odot}$ & Stellar luminosity\tablenotemark{a}  \\
$M_\star$ & 
$M_{\odot}$ & Stellar mass\tablenotemark{b}  \\
$e$   & $0.00$ & Orbital eccentricity   \\
$\omega$   & $0.00$ & Argument of the pericenter   \\
$\epsilon$ & $23.44$ & Axis obliquity  \\
$g$ & 9.8 m s$^{-2}$ & Surface gravitational acceleration\tablenotemark{c} \\
$P_\text{rot}$ & 1 d & Rotation period \\
$f_o$  & $0.70 $ & Ocean fraction (constant in all latitude zones) \\ 
\enddata
\tablenotetext{a}
{The solar luminosity $L_{\odot}$ is calculated from the adopted value of solar constant, $q_0$ (Table \ref{tabEarthData})} 
\tablenotetext{b}
{The stellar mass, in conjunction with the semi-major axis $a$, determines the 
orbital period adopted in each simulation.}
\tablenotetext{c}
{The surface gravitational acceleration is used in the radiative calculations of the OLR (Section \ref{sectOLRcal})}
\label{multirunPar}
\end{deluxetable}

\end{document}